\documentclass[aps,twocolumn,superscriptaddress,nofootinbib,longbibliography]{revtex4-1}

\usepackage{amsmath,amsfonts,amssymb}
\usepackage{mathtools}
\usepackage{wrapfig}
\usepackage{braket}
\usepackage{graphicx}
\usepackage{bbm}
\usepackage{graphics}
\usepackage{floatrow}
\usepackage[caption=false]{subfig}
\usepackage{setspace}
\usepackage{array}
\usepackage{enumitem}
\usepackage{sidecap}
\usepackage{xcolor}

\usepackage{xr-hyper}
\usepackage{hyperref}

\sidecaptionvpos{figure}{t}

\newtheorem{theorem}{Theorem}

\newcommand{\mb}[1]{\mbox{\boldmath$#1$}}

\begin{document}

\title{Using an atom interferometer to infer gravitational entanglement generation}

\author{Daniel Carney}
\affiliation{Joint Center for Quantum Information and Computer Science/Joint Quantum Institute, 
University of Maryland/NIST, College Park, MD 20742, USA}
\affiliation{Fermi National Accelerator Laboratory, Batavia, IL 60510, USA}
\author{Holger M\"{u}ller}
\affiliation{Department of Physics, University of California, Berkeley, CA 94720, USA}
\author{Jacob M. Taylor}
\affiliation{Joint Center for Quantum Information and Computer Science/Joint Quantum Institute, 
University of Maryland/NIST, College Park, MD 20742, USA}
\date{\today}

\begin{abstract}
If gravitational perturbations are quantized into gravitons in analogy with the electromagnetic field and photons, the resulting graviton interactions should lead to an entangling interaction between massive objects. We suggest a test of this prediction. To do this, we introduce the concept of interactive quantum information sensing. This novel sensing protocol is tailored to provable verification of weak dynamical entanglement generation between a pair of systems. We show that this protocol is highly robust to typical thermal noise sources. The sensitivity can moreover be increased both using an initial thermal state and/or an initial phase of entangling via a non-gravitational interaction. We outline a concrete implementation testing the ability of the gravitational field to generate entanglement between an atomic interferometer and mechanical oscillator. Preliminary numerical estimates suggest that near-term devices could feasibly be used to perform the experiment.

\end{abstract}

\maketitle

\section{Introduction}

If a particle is in a superposition of two locations, will its gravitational field also be in a superposition, and can this field generate entanglement with another system? This foundational question  \cite{feynman1971lectures,page1981indirect} has received considerable attention \cite{diosi1984gravitation,penrose1996gravity,kafri2015bounds,bose2017spin,Marletto:2017kzi,Haine:2018bwu,Chevalier:2020uvv,Howl:2020isj,Anastopoulos:2020cdp,Matsumura:2020law,carney2019tabletop}. Proposed experimental tests to detect entanglement due to gravity based on Bell tests (or more generally, entanglement witnesses \cite{HORODECKI19961,terhal2000bell}) require performing measurements on both subsystems and are challenging in practice. As a result, there is still no direct experimental evidence as to whether gravitational interactions generate entanglement. Here, we propose a test that only requires observing a single subsystem \cite{feynman1963theory,caldeira1983path,joos1985emergence,zurek2003decoherence}. We show that, if an interaction (such as gravity) between two systems can cause both decoherence (collapse) and recoherence (revival) of a subsystem, then for restricted classes of systems the interaction is \emph{necessarily} capable of generating entanglement. We propose a concrete implementation based on atom interferometry \cite{kasevich1992measurement,santarelli1999quantum,gross2010nonlinear,xu2019probing}, in which an atom in a superpositon of being in one of two interferometer arms interacts with a low-frequency mechanical resonator \cite{lee2020new,catano2020high}; the signal for entanglement-generation is a collapse and revival of the atomic interference fringes due to the periodic motion of the resonator. The experiment does not require  preparing a non-classical state of the oscillator and can in fact be enhanced by placing the oscillator in a thermal state, which appears to make this experiment feasible with near-term devices. 

The relation of such an experiment to the quantization gravity is a subject of intense current study \cite{belenchia2018quantum,Christodoulou:2018cmk,Marshman:2019sne,Galley:2020qsf}. These experiments operate in a regime where the energy density (or equivalently, spacetime curvature), is far below the Planck scale $\rho \ll m_{\rm Pl}/\ell_{\rm Pl}^3 \sim 10^{123}~{\rm eV}/{\rm cm^3}$. Thus the non-linearity of the gravitational interaction is very weak, and one can treat the metric $g_{\mu\nu}$ as a linear perturbation around flat spacetime. In this limit, one can quantize the gravitational perturbations (``gravitons'') in exact analogy with quantum electrodynamics; graviton exchange generates a two-body Newton potential operator
\begin{equation}
\label{VN}
V_N = -\frac{G_N m_1 m_2}{|\mb{x}_1-\mb{x}_2|}
\end{equation}
between a pair of masses, just as photons generate the Coulomb potential \cite{Feynman:1963ax,t1974one,deser1974one,Veltman:1975vx,donoghue1994general,burgess2004quantum}. We review some standard demonstrations of this in appendix \ref{appendix-graviton}. In equation \eqref{VN}, $\mb{x}_{1,2}$ are the position \emph{operators} on a pair of masses, and thus this interaction can generate entanglement. However, there are dissenting opinions \cite{penrose1996gravity,howl2019exploring,Tilloy:2019hxe,Bruschi:2020xbm} about whether gravity should be quantized in this way, and indeed one can produce models where classical gravitational interactions can arise but without generating entanglement \cite{Kibble:1979jn,kafri2014classical,Oppenheim:2018igd,Kent:2020gov}, providing substantial motivation to perform tests of \eqref{VN}.

\begin{figure*}
    \includegraphics[width=0.95\linewidth]{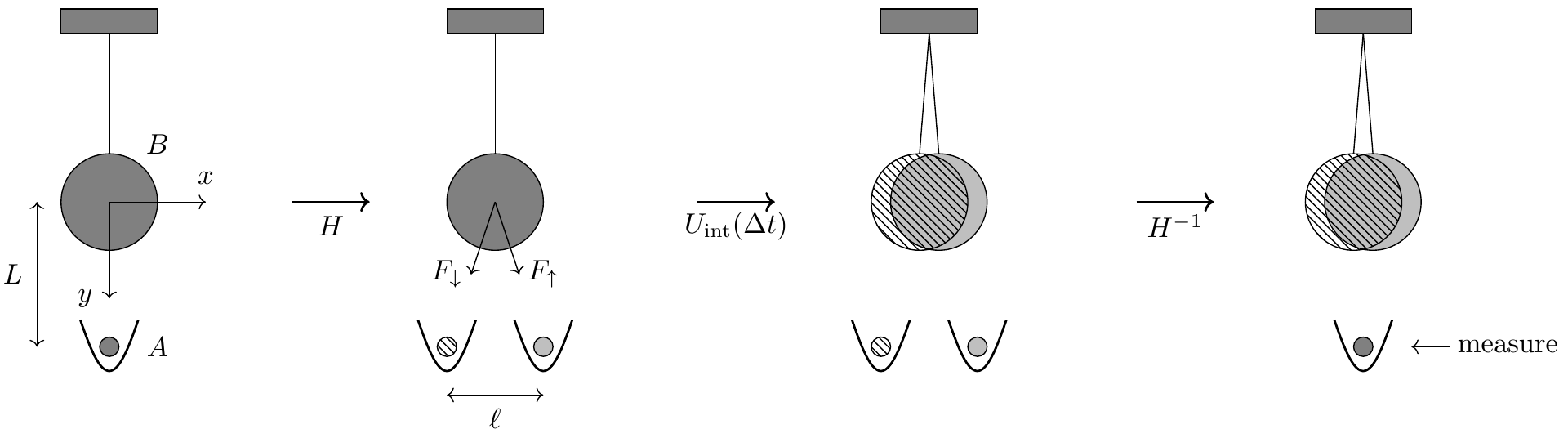} 
    \caption{Implementation of the basic protocol using an atom interferometer and a suspended pendulum (see Section \ref{section-experiment}). A trapped atom (labeled A) is prepared some distance $L$ away from a mechanical resonator (B, here pictured as a pendulum). The atom is then put into a superposition of two different locations separated by $\ell$, effecting a Hadamard gate $H$. This generates a state-dependent force between the atoms and resonator, leading to motion in opposite directions for some time $\Delta t$. Finally, the atom state is recombined using the inverse Hadamard gate and measured to check for decoherence caused by the atom-mechanical interaction. When the resonator undergoes a complete period of motion, its state no longer depends upon the atoms and coherence is recovered for the interferometer.}
    \label{fig:setup}
\end{figure*}


The ability to test such a weak entanglement signal relies entirely on our central technical result, a novel sensing protocol which we refer to as \emph{interactive quantum information sensing}. This is a detection scheme tailored specifically to the verification of weak dynamical entanglement generation. The traditional methods to detect entanglement in bipartite systems $H_A \otimes H_B$ use non-local measurements \cite{HORODECKI19961,terhal2000bell}, and can be very difficult in practice with noisy systems and weak entanglement. However, in the past two decades, more sophisticated methods have been developed to address these types of problems  \cite{guhne2009entanglement,pezze2018quantum}. We suggest here a new protocol which relies on time-dependent measurements on a single subsystem. Within standard quantum mechanics, system $A$ will decohere---evolve from a pure to mixed state---if it becomes entangled with another system $B$ which is not measured \cite{feynman1963theory,caldeira1983path,joos1985emergence,zurek2003decoherence}. This loss of coherence can be observed via an interference measurement on $A$ alone. Simple decoherence could be explained by entanglement but also by, for example, random classical noise \cite{stern1990phase}. However, if the same interaction can cause both decoherence and recoherence of $A$, in a manner controlled by $B$, then for certain classes of systems we prove that the interaction is necessarily capable of generating entanglement between subsystems $A$ and $B$. This protocol provides an indirect test of the quantum communication capabilities of the two systems, and is a limited probe of the family of quantum channels associated with the interaction between the two systems. The interplay between the information theoretic channel properties and the physical interaction provides our suggested nomenclature.

We outline the interactive sensing protocol in sections \ref{section-revival} and \ref{section-theorem}. We find the remarkable result that using an initial state at high temperature can \emph{increase} the sensitivity of the protocol, because it can increase the rate of entanglement generation and lead to a thermally-enhanced collapse and revival signal. In section \ref{section-noise} we demonstrate that this conclusion is robust to typical sources of noise, essentially because the test does not involve producing large superpositions of the non-observed subsystem. In section \ref{section-boosted} we show how to further enhance the protocol using pre-entangled initial conditions. Finally, we outline an experimental realization with gravitational entanglement generation between an atom interferometer and a mechanical oscillator in section \ref{section-experiment}, before concluding with a discussion of implications and loopholes in section \ref{section-conclusions}.

\section{Collapse and revival dynamics}

\label{section-revival}

To begin, we illustrate the basic idea of the collapse-and-revival dynamics with an example. The setup is similar to electron spin echo envelope modulation \cite{rowan1965electron,dikanov1992electron} and the cavity QED experiments of Haroche \emph{et al.} \cite{raimond2001manipulating}. Consider an harmonic oscillator $B$ coupled to a two-state system $A$ through the Hamiltonian
\begin{equation}
\label{hamiltonian}
H = \omega a^{\dagger} a + g (a+a^{\dagger}) \sigma_z.
\end{equation}
In section \ref{section-experiment} we give an implementation of this Hamiltonian where the oscillator $B$ is a mechanical resonator, the two-state system $A$ corresponds to an atom located in one of two spatial locations, and $g \ll \omega$ is set by the atom-oscillator gravitational interaction \eqref{VN}, so $g$ is proportional to Newton's constant $G_N$. The essential idea is to do an interferometry measurement on the two-state system $A$ (the ``control'') in the presence of system $B$ (the ``target''). The key is the dynamical response of the target system $B$ to a superposition of $A$.
    
\begin{figure*}
\subfloat{
\begin{tabular}[b]{c}
    \includegraphics[width=.35\linewidth]{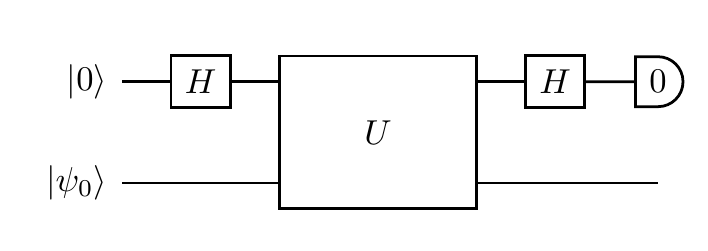} \\
    \includegraphics[width=.35\linewidth]{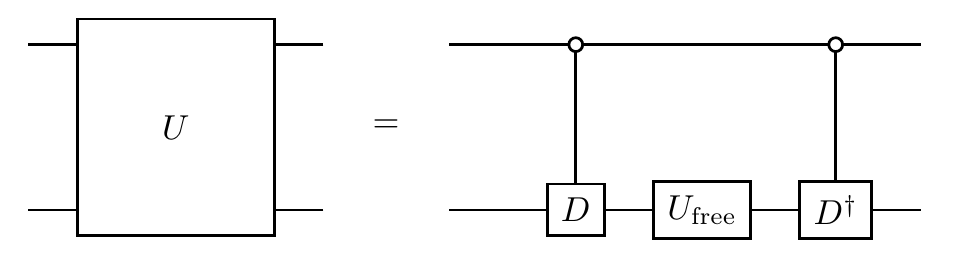}
    \end{tabular}}
\subfloat{
\centering
\includegraphics[width=.3\linewidth]{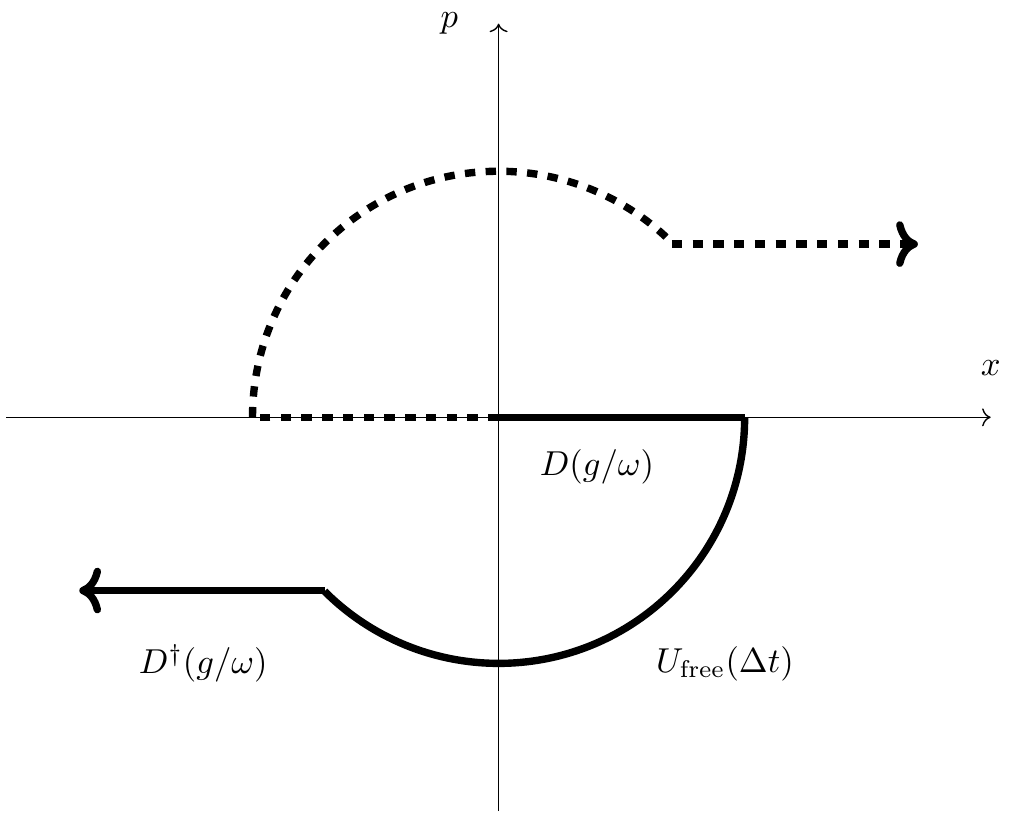}}
\subfloat{\includegraphics[width=.3\linewidth]{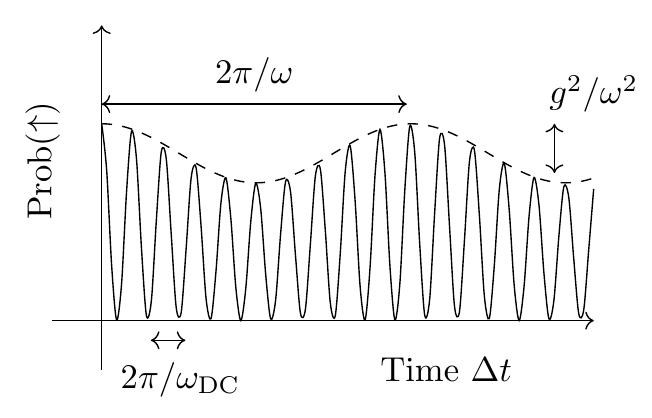}}
\caption{Equivalent circuit (left) and phase space (center) descriptions of the experiment, and schematic interferometric data output (right). In the circuit, the top line represents the atom and bottom line the resonator. The large box represents joint evolution of the trapped atom and the resonator, which can be decomposed into conditional displacement of the resonator, followed by free evolution and an inverse displacement operator. This sequence can be visualized in the phase space of the oscillator, where the solid and dashed lines represent the two oscillator evolutions conditioned on the two possible atomic locations. Interferometric measurement of the atom population will show rapid fringes with frequency $\omega_{\rm DC}$ due to any stray DC accelerations (e.g., due to electric fields, Earth's gravity, or off-center location of the resonator or atom), modulated by an overall reduction and then increase due to the atom-resonator entanglement. Resonator motion over a full period leads to nominal full recovery of the fringes.}
\label{fig:setup2}
\end{figure*}

To understand the entanglement dynamics generated by \eqref{hamiltonian}, it is useful to note that the time evolution operator can be re-written
\begin{equation}
\label{Uop}
U(t) = e^{-i H t} = D^{\dagger}\left(\sigma_z \lambda \right) e^{-i \omega a^{\dagger} a t} D\left(\sigma_z \lambda \right)
\end{equation}
up to an overall phase, where
\begin{equation}
D(\alpha) \equiv \exp \left\{ \alpha a^\dag - \alpha^* a \right\}
\end{equation}
is the usual displacement operator.\footnote{To see this, note that $D^{\dagger}(\alpha) a^{\dagger} a D(\alpha) = | a + \alpha|^2$ and expand the free evolution operator $e^{-i \omega a^{\dagger} a t}$ in the middle of \eqref{Uop}.} Here and throughout, we will use the dimensionless quantity
\begin{equation}
\lambda \equiv \frac{g}{\omega}.
\end{equation}
This is the length, measured in units of the zero-point length $x_0$, that the oscillator equilibrium is displaced under the force from the atom. This ratio will set the scale of all observables considered in this paper. 

Observing the collapse-and-revival can be done with a typical interferometric measurement. Consider starting the full system in its decoupled ground state $\ket{0}_A \otimes \ket{0}_B$. The interferometry experiment then proceeds by performing a Hadamard gate (or any other beamsplitter operation) on the two-state system $A$, $\ket{0} \to (\ket{0} + \ket{1})/\sqrt{2}$, evolving the joint system for some time $t$, performing the inverse Hadamard gate to recombine the two-level system, and then measuring its population. Mathematically, this proceeds as follows:
\begin{align}
\begin{split}
\label{toyevolution}
\ket{\psi} & = \ket{0}_A \otimes \ket{0}_B \\
& \xrightarrow{H} \frac{\ket{0}_A + \ket{1}_A}{\sqrt{2}} \otimes \ket{0}_B \\
& \xrightarrow{U_{\rm int}} \frac{\ket{0}_A \ket{\delta}_B + \ket{1}_A \ket{-\delta}_B}{\sqrt{2}} \\
& \xrightarrow{H^{\dagger}} \ket{0}_A \frac{\ket{\delta}_B + \ket{-\delta}_B}{2} + \ket{1}_A \frac{\ket{\delta}_B - \ket{-\delta}_B}{2}.
\end{split}
\end{align}
Here, the conditionally-evolved states of the oscillator are simply coherent states
\begin{equation}
\label{coherentstate}
\ket{\pm \delta}_B = D\left(\pm \lambda (e^{-i \omega t} -1)\right) \ket{0}.
\end{equation}
If we now measure the two-state system $A$, we find for example that the probability of being in the $\ket{0}$ state is
\begin{equation}
\label{decoherence}
P_A(0) = \frac{1}{2} + \frac{1}{2} {\rm Re} \braket{ \delta | -\delta }_B = \frac{1}{2} \left( 1 + e^{-8 \lambda^2 \sin^2(\omega t/2)} \right).
\end{equation}
We see that the interference term is reduced, with a period set by the oscillator frequency $\omega$. In particular, at half-period we have a maximum reduction of the phase contrast, and after a full period the contrast is completely restored, as in Fig. \ref{fig:setup2}.

Before moving on, we mention for later use an alternative calculation of the same effect. Consider the Pauli lowering operator $\sigma_- = (\sigma_x - i \sigma_y)/2$ on the two-level system. The expectation value $\braket{\sigma_-(t)}$ tracks the loss of phase contrast; we will refer to the absolute value as the interferometric visibility $V = |\braket{\sigma_-}|$. Using the time-evolution operator \eqref{Uop}, we have
\begin{align}
\begin{split}
\label{visibility}
\sigma_-(t) & = U^{\dagger}(t) \sigma_- U(t) \\ 
& = D^\dag(-\lambda) e^{i \omega a^{\dagger} a t} D(-\lambda)  \sigma_- D^\dag(\lambda) e^{-i \omega a^{\dagger} a t} D(\lambda) \\
& = \sigma_- D(2 \lambda (1-e^{i \omega t})).
\end{split}
\end{align}
This is easy to show by working with explicit components in the $\sigma_z$ basis, where $\sigma_- = \ket{1} \bra{0}$. With an oscillator initially in the ground state, this gives
\begin{equation}
\label{sigma-}
\braket{\sigma_-(\pi/\omega)} = \braket{\sigma_-(0)} e^{-8\lambda^2}, \ \ \ \braket{\sigma_-(2\pi/\omega)} = \braket{\sigma_-(0)}.
\end{equation}
Here we see again the loss of phase contrast at half period followed by the revival at a full period. 

Up to this point, we have assumed that the oscillator was initialized in its ground state $\ket{0}$. In a realistic implementation--particularly one where the oscillator is a massive mechanical object--the oscillator will instead start in a mixed state, such as a thermal state, due to its coupling to an environment. Although one may be concerned that this would destroy the revival of coherence in the atom, it turns out that not only does the revival persist, but in fact the relative contrast between decoherence and revival is \emph{enhanced} so long as the thermalization time scale remains very long. That the revival persists is a consequence of the harmonic potential: after a full-period, the state of the oscillator must return to its initial condition.

To see this, consider first the oscillator initialized to an arbitrary coherent state $\ket{\alpha}$. Using \eqref{visibility}, we have
\begin{align}
\begin{split}
\braket{\alpha | \sigma_-(t) | \alpha} & = e^{-2\lambda [\alpha^*(1-e^{i \omega t}) - \alpha(1-e^{-i \omega t})]} \\
& \times e^{-8 \lambda^2 \sin^2 (\omega t/2)} \braket{\sigma_-(0)}.
\end{split}
\end{align}
We see the complete revival after a full period, while at half period we now pick up a phase involving the initial oscillator momentum $p_{\alpha} = \alpha+\alpha^*$. To obtain the thermal-state result, one can now average over the coherent states (i.e. use the oscillator density matrix $\rho_{\rm th} = \int d^2\alpha e^{-|\alpha|^2/\bar{n}}/(\pi \bar{n}) \ket{\alpha} \bra{\alpha}$, with $\bar{n}$ the thermal phonon occupancy). The result for the qubit visibility is
\begin{equation}
\label{thermalrevival}
V_{\rm th}(t) = \exp \left[-8 \lambda^2 (2 \bar{n}+1) \sin^2(\omega t/2) \right].
\end{equation}
In particular, we have $V_{\rm th}(2\pi/\omega) = 1$, showing a full revival of the qubit coherence after a full oscillator period. On the other hand, at half-period, we have $V_{\rm th}(\pi/\omega) = \exp\left[ -8\lambda^2 \left(2 \bar{n} + 1 \right) \right]$, an enhancement to the loss of visibility by a factor of $\bar{n}$. Thus, starting with a thermal state increases the contrast between the `dip' of coherence halfway through oscillation and the recovery at full oscillation. The experiment is  \emph{easier} with a hot oscillator.

\section{Revival verifies entanglement generation}
\label{section-theorem}

As this example clearly shows, entanglement generation between two systems $A$ and $B$ can cause periodic collapse and revival of $A$'s wavefunction. The crucial question is then: does observation of this collapse and revival  \emph{necessarily} require  entanglement generation between $A$ and $B$? Our central result says that the answer is yes, under some particular assumptions. We characterize this with a theorem:

\begin{theorem} \label{theorem1} Let $L$ be a channel on $H_A \otimes H_B$, where $H_A$ is a two-state system and $H_B$ is arbitrary. Assume that
\begin{enumerate}[label=(\alph*)]
    \item \label{semigroup} The channel $L$ generates time evolution, in a manner consistent with time translation invariance, thus obeying a semigroup composition law $L_{t \to t''} = L_{t \to t'} L_{t' \to t''}$ for all $t \leq t' \leq t''$,
    \item \label{population} The two-level subsystem $H_A$ has its populations preserved under the time evolution, $\sigma_z(t) = \sigma_z(0)$, and
    \item \label{separable} $L$ is a separable channel \cite{rains1999rigorous}: all of its Krauss operators are simple products. In particular, this means that any initial separable (non-entangled) state evolves to a separable state: $\rho(t) = L_t[\rho(0)]$ is separable for all separable initial states $\rho(0)$.  
\end{enumerate}
Then the visibility $V(t) = |\braket{\sigma_-(t)}|$ is a monotonic function of time.
\end{theorem}

Here, we have modeled the time evolution of the $A$-$B$ system as a quantum channel $L$, a map on density matrices $\rho(t) = L_t[\rho(0)]$. For example, within standard quantum mechanics, the unitary evolution of the universe ($A,B$ and their environment $C$, including the experimentalist) generates such a channel for the reduced $A-B$ evolution. Suppose that we can experimentally convince ourselves that time-translation invariance in the form \ref{semigroup} and population condition \ref{population} hold. Then the theorem says that if $L$ cannot generate entanglement \ref{separable}, then the only possible evolution for the qubit $A$ is to have its interferometric visibility decay monotonically. Thus if we observe non-monotonic visibility like the oscillatory signal described above, we can conclude that the channel must be capable of generating entanglement.

We note that non-entangling channels still allow for non-trivial interactions. For example, semiclassical gravity $G_{\mu\nu} = 8\pi \braket{T_{\mu\nu}}$ (appropriately completed by a modified version of the Schr\"{o}dinger equation) is of this form \cite{kafri2015bounds}. On the other hand, the graviton model will produce an entangling channel. 
 
We now give a proof of this theorem. By assumption \ref{semigroup}, there exists a generator $\mathcal{L}$ of $L_t$ of Lindblad form \cite{lindblad1976generators,gorini1976completely}:
\begin{equation}
\label{lindblad}
    \dot{\rho} = \mathcal{L} \rho = -i[H,\rho ] - \sum_{j} \gamma_j \left[E_j^\dag E_j \rho + \rho E_j^\dag E_j - 2 E_j \rho E_j^\dag \right].
\end{equation}
These Lindblad operators $E_j$ are highly constrained by the separability assumption, because they cannot be used to generate $A-B$ entanglement. To make this precise, we write the channel in its Krauss representation $L[\rho] = \sum_{j \geq 0} L_j \rho L_j^\dagger$. Expanding for small times and comparing to \eqref{lindblad}, one finds that the Krauss operators $L_j$ take the form, to lowest order in $dt$, 
\begin{equation}
L_0 = 1 - i H dt + K dt, \ \ L_j = E_j \sqrt{dt}, \ \ K = -\frac{1}{2} \sum_{j>0} E_j^\dagger E_j.
\end{equation}
See, for example, chapter 3 of \cite{preskill1998lecture}. Now we invoke the separability criterion \ref{separable}, which says that the Krauss operators for $j > 0$ take the form of simple product operators, i.e. $E_j = A_j \otimes B_j$ \cite{rains1999rigorous}. Furthermore, the separability of $L_0$ to order $dt$ means that $L_0 = (1 + A_0 dt) \otimes (1 + B_0 dt)$ for some $A_0,B_0$, and this can only be satisfied if both $H$ and $E_j^\dag E_j$ can be written as sums of operators acting either on $\mathcal{H}_A$ \emph{or} $\mathcal{H}_B$. This in turn requires that for each $j > 0$, either $A_j^\dag A_j = 1_A$ or $B_j^\dag B_j = 1_B$. Finally, we impose the requirement \ref{population} that the atom populations are invariant. This means that $\dot{\sigma}_z = 0$. The only possible non-trivial interaction term which satisfies these requirements is $E_z = \sigma_z \otimes B$, with $B$ any operator on $H_B$. 

We are then left with the the very simple form of the Lindblad generator:
\begin{equation}
\label{lindblad-simple}
    \mathcal{L}\rho = - \gamma \left[ B^\dag B \rho + \rho B^\dag B - 2 B \sigma_z \rho \sigma_z B^\dag \right] + \mathcal{L}_A + \mathcal{L}_B.
\end{equation}
Here $\mathcal{L}_{A(B)}$ are Lindblad operators (including Hamiltonians) acting only on $\mathcal{H}_{A(B)}$, and $\mathcal{L}_{A}(\sigma_z) = 0$. With this result for the channel's structure, we can compute the time derivative of the interferometric visibility $V(t) = |\braket{ \sigma_-(t)}|$. Since $[H,\sigma_z] = 0$, the most general qubit Hamiltonian is a sum of $\sigma_z$ and the identity. We thus have, in the Heisenberg picture,
\begin{align}
    \begin{split}
        \Braket{  \frac{d\sigma_-}{dt}} & = -i \braket{[H,\sigma_-]} + \gamma \left[ \braket{ E_z^{\dagger} \sigma_- E_z } - \frac{1}{2} \braket{ \left\{ E_z^{\dagger} E_z, \sigma_- \right\}} \right]\\
        & = 2 (-i \omega_0 -\gamma) \braket{\sigma_-},
    \end{split}
\end{align}
where the oscillatory term is generated by the qubit Hamiltonian. Taking the absolute value to compute the visibility $V = |\braket{\sigma_-}|$ removes the oscillating phase and we have 
\begin{equation}
\label{visibility-answer}
\frac{dV}{dt} = - 2 \gamma V,
\end{equation}
so it is monotonically decreasing, as we set out to prove.

\section{Effects of noise during evolution}

\label{section-noise}

The sensing protocol is subject to errors caused by random noise during the time evolution. In a typical realization, the dominant sources of this continuous noise will consist of thermal load on the oscillator and dephasing in the atomic system (from, e.g., background fields and gas weakly measuring the atomic position \cite{joos1985emergence,gallis1990environmental}). These sources of noise can be modeled by a Lindblad evolution of the form
\begin{equation}
\label{lindblad-noise}
\dot{\rho} = -i[H,\rho] - \sum_{i} \frac{1}{2} \{ L_i^{\dagger} L_i, \rho \} - L_i \rho L_i^{\dagger},
\end{equation}
where the error operators are $L_i \in \{ \sqrt{\bar{n} \gamma_m} a^{\dagger}, \sqrt{(\bar{n}+1) \gamma_m} a, \sqrt{\gamma_a} \sigma_z \}$. The decay rates of the oscillator and atom are $\gamma_m,\gamma_a$, respectively, and $\bar{n}$ is the thermal phonon occupancy. This description should be accurate for times similar to or shorter than the damping time $1/\gamma_m$, and assuming only small changes over time in the mechanical frequency.

It is possible to analytically solve for the atomic visibility $\eqref{visibility}$ in the presence of this noise, using an explicit Ohmic heating model where the bath is taken to be an infinite set of bosonic modes linearly coupled to the mechanical system. The same displacement-operator picture used in \eqref{Uop} generalizes to this linear bath (see appendix \ref{appendix-noise}). One finds that the visibility at half and full-period evolution is given by
\begin{align}
\begin{split}
\label{SNR}
V(\pi/\omega) & = \exp[-\pi \gamma_a/\omega] \exp [-8 \lambda^2 (2\bar{n}+1)] \\
V(2 \pi/\omega) & = \exp[-2 \pi \gamma_a/\omega] \exp [-8 \lambda^2 (2\bar{n}+1)/Q].
\end{split}
\end{align}
Here we have assumed the mechanical damping factor $Q = \omega/\gamma_m \gg 1$. 

This recovers the previous result for the visibility \eqref{thermalrevival}, up to an overall exponential damping from the atomic dephasing and small correction from mechanical heating. Neglecting atomic dephasing, the visibility at half period is exactly the same as \eqref{thermalrevival}, while at full period, for $Q \gg 1 \gg \bar{n} \lambda^2$, we have $V(2\pi/\omega) \approx 1$, i.e. we have full recovery up to a correction at order $1/Q$. Thus, with a sufficiently high-$Q$ oscillator, and with atomic coherence times longer than the mechanical period $\gamma_a \lesssim \omega$, damping does not pose a substantial barrier to the experiment.

Before moving on, we consider the effects of decoherence from another inevitable source: blackbody radiation of the oscillator. Here we are discussing position superpositions of the oscillator at distances of about $\lambda x_0$. With the sorts of experimental parameters we suggest later, this will be a length many orders of magnitude smaller than a typical blackbody photon wavelength (or ambient gas molecule's de Broglie wavelength). Thus these interactions will be incapable of efficiently decohering the oscillator, because they are too long-wavelength to efficiently measure the oscillator's position \cite{gallis1990environmental}.

\section{Protocol linear in the weak coupling}

\label{section-boosted}

\begin{figure*}
\includegraphics[width=.35\linewidth]{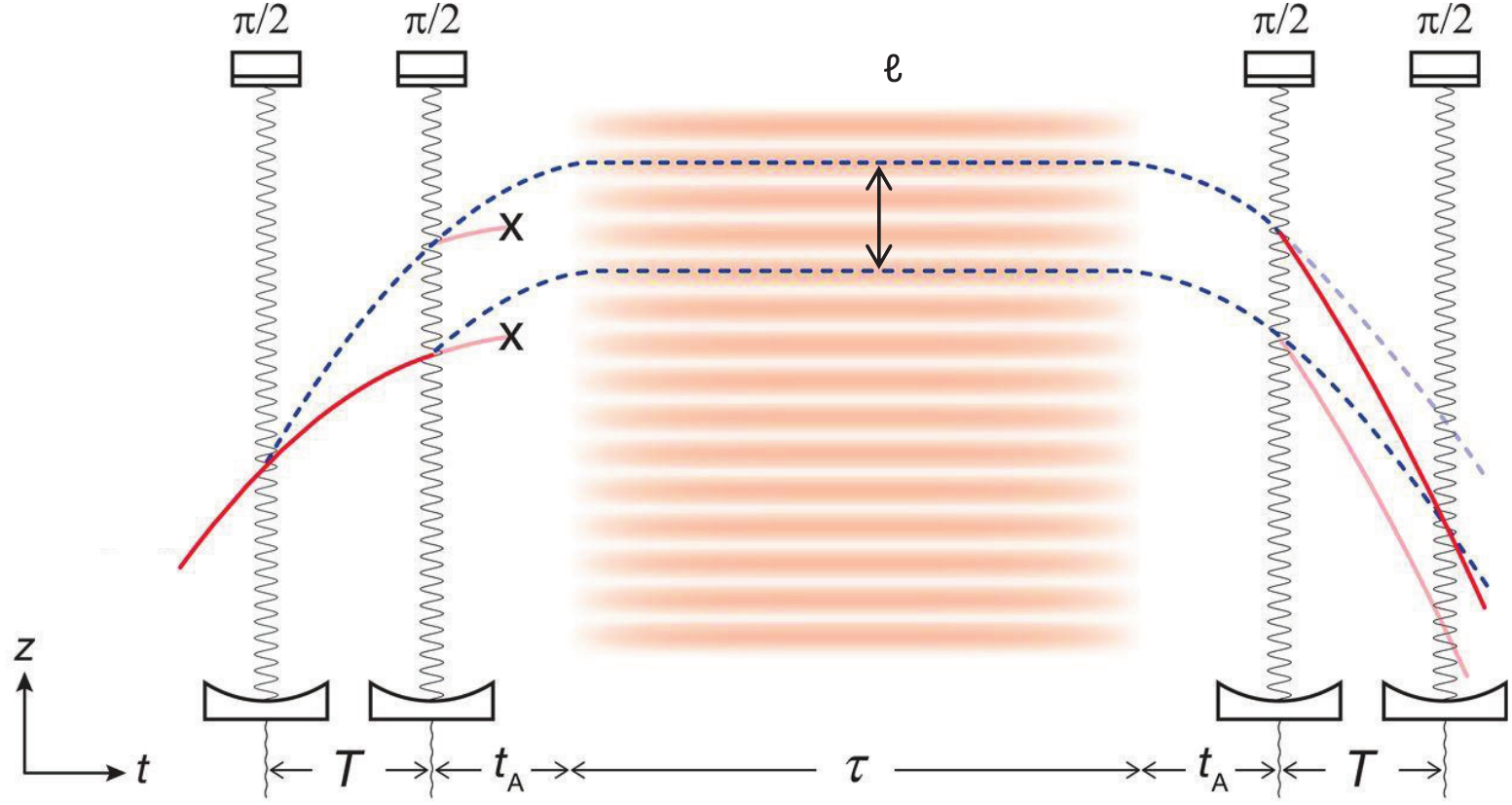} \includegraphics[width=0.63\linewidth]{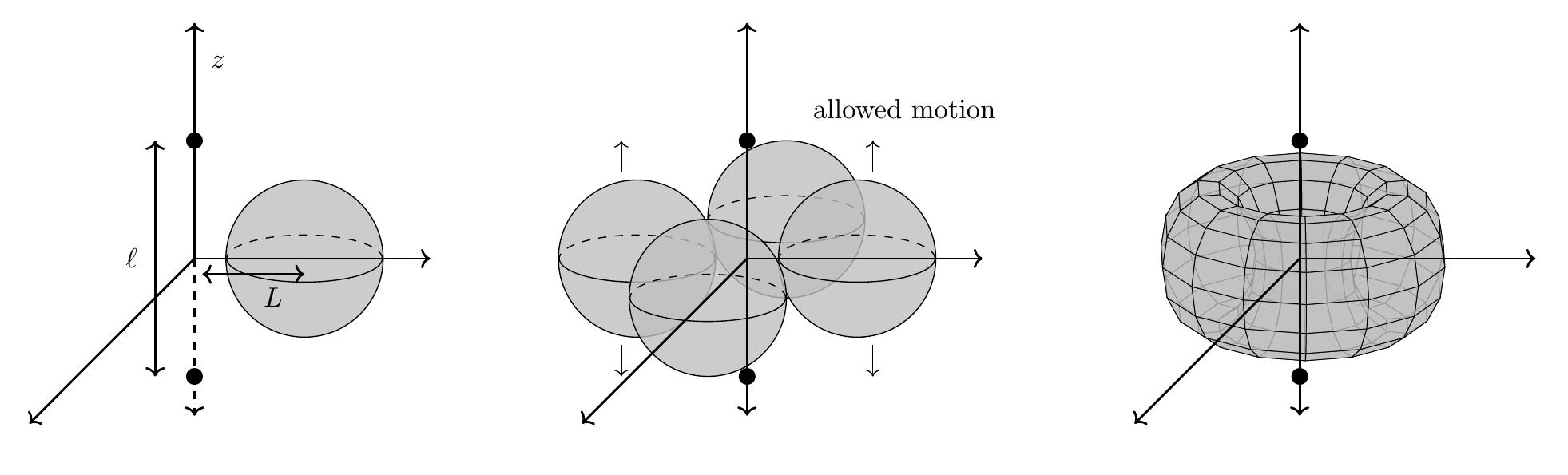}
\caption{Left: Experimental realization of the atomic system as a lattice interferometer. The lines marked ``x" denote populations  that do not interfere.  Right: Some example implementations with one or more mechanical masses connected rigidly. Small black dots represent the atom. In each case, the mechanical system is restricted to oscillate along the $z$-axis. More masses enable a stronger gravitational coupling. A natural limiting case would be to use a toroidal mass. In the example with a single sphere, we have $R = \sqrt{L^2 + (\ell/2)^2}$ and $\kappa = 1$.}
\label{fig:setup3}
\end{figure*}

Our basic observable \eqref{decoherence} is quadratic in the ratio $\lambda = g/\omega$, which for a weak coupling is a small dimensionless number. Here we suggest a ``boosted'' method in which linear sensitivity can be achieved by first preparing an \emph{entangled} state of the atom-oscillator system (as demonstrated, for example, in \cite{karg2020light,thomas2021entanglement}).

Let $\lambda' = g'/\omega$, where $g$ is the coupling of interest (e.g., gravity) and $g'$ is some other coupling. Consider performing a $\pi$ gate with the coupling $V_{\rm int} = (g + g') \sigma_z x$. This will produce an initial entanglement set by displacement operators $D(\pm(\lambda+\lambda'))$, as in equation \eqref{visibility}. Turning off the non-gravitational $g'$ coupling then leads to only a partial revival of the atomic signal at later times $t > \pi/\omega$. This leads to the visibility, for $t > \pi/\omega$,
\begin{equation}
\label{signal-boosted-timeseries}
V_{\rm b}(t) = \exp\left[ -8(2\bar{n}+1) \left(  \lambda'^2 + 2 \lambda \lambda' \sin^2 \frac{\omega t}{2} + \lambda^2 \sin^2 \frac{\omega t}{2} \right) \right].
\end{equation}
A detailed calculation is given in appendix \eqref{appendix-boostedcalc}. For times $0 < t < \pi/\omega$, the visibility is given by the previous result \eqref{thermalrevival} but with $\lambda \to \lambda + \lambda'$. 

The observable we are interested in is the difference in visibility at half-period and full-period:
\begin{align}
\begin{split}
\label{SNR-thermalboosted}
\Delta V_{\rm b} & = V_b(2\pi/\omega) - V_b(\pi/\omega) \\
& \approx \exp[-8(2\bar{n} + 1) \lambda'^2](1 - 16(2\bar{n} + 1) \lambda' \lambda + O(\lambda^2)),
\end{split}
\end{align}
assuming $\lambda \ll \lambda'$. We see again that using an initially ``hot'' resonator increases the relative visibility. However, here the observable is \emph{linear} in the weak gravitational coupling $\lambda$. We note that if $\bar{n}$ or $\lambda'$ are too large, the signal will be destroyed by the overall prefactor $e^{-8 (2\bar{n} + 1) \lambda'^2}$. The optimal solution is to tune the non-gravitational coupling to satisfy $\lambda'_{\rm opt} = 1/\sqrt{8(2\bar{n} + 1)}$,
in which case the prefactor is order one, and the relative visibility is given roughly by $\Delta V_b \approx \sqrt{8(2\bar{n} + 1)} \lambda$. Use of this boosted protocol substantially improves the viability of an experiment with a weak coupling $g$. We note that this protocol does not violate our assumptions about time-translation invariance in Theorem \ref{theorem1}: once the extra $g'$ coupling is turned off, the entire system proceeds in a time-independent fashion.

\section{Experimental implementation with atom interferometry}
\label{section-experiment}

\begin{SCfigure*}
\includegraphics[width=.7\linewidth]{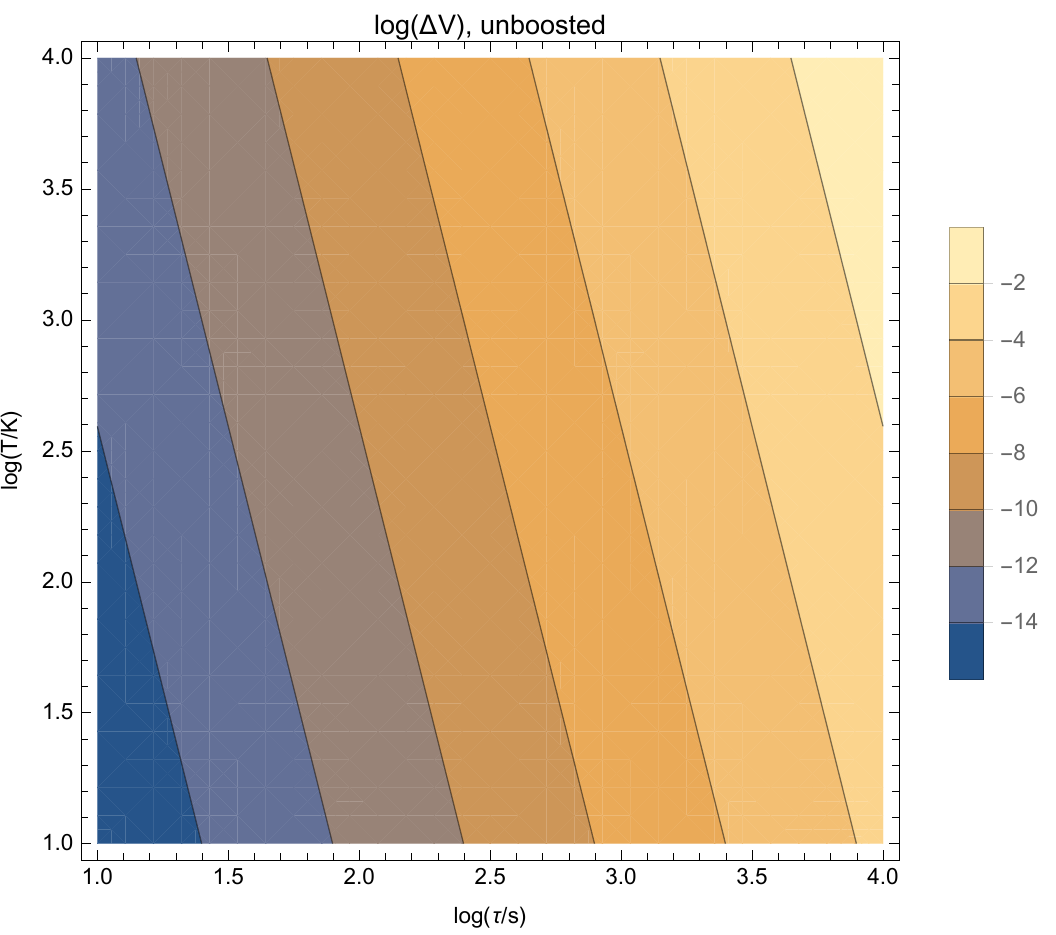}
\includegraphics[width=.7\linewidth]{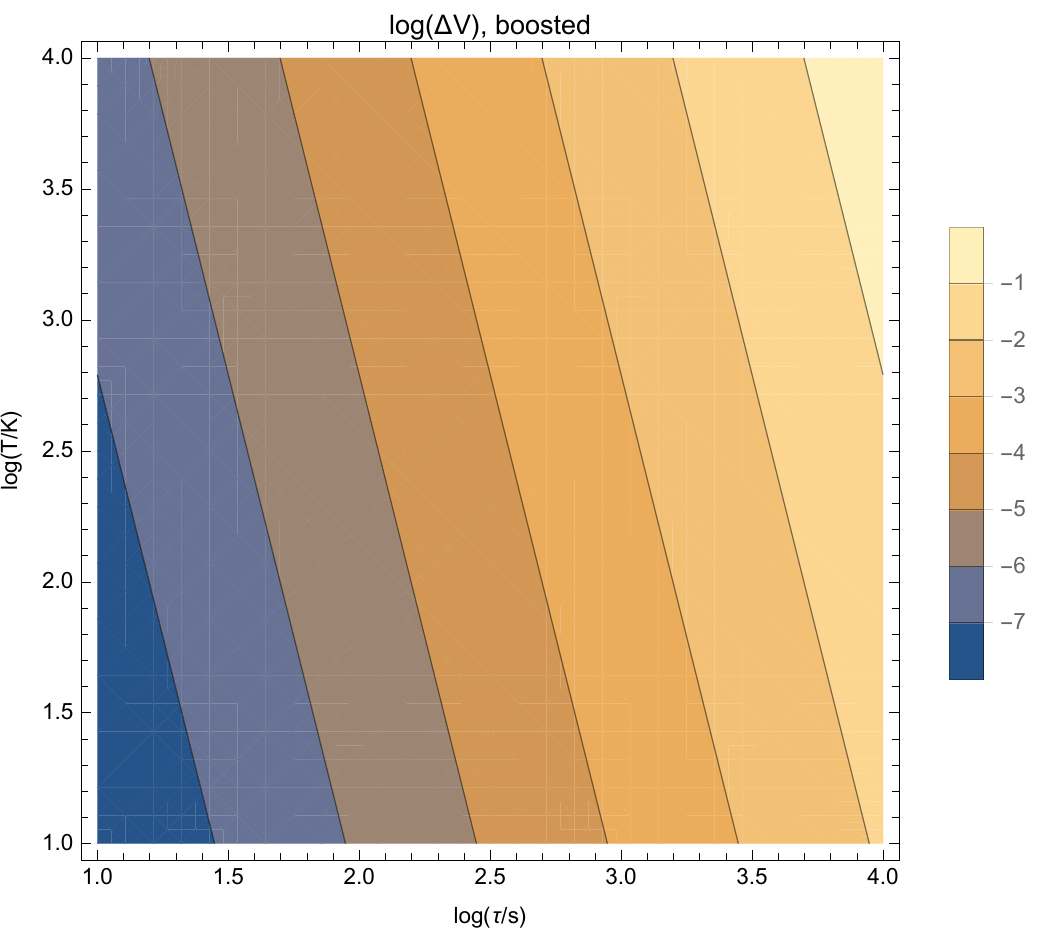} \ \ \ \ \label{fig:DeltaV} \caption{\protect\rule{0ex}{5ex}Logarithm $\log_{10}$ of the visibility change $\Delta V$ as function of the logarithms of the hold time $\tau$ in seconds and the temperature $T$ in Kelvin for the unboosted scheme (left) and the boosted scheme (right). The plots assume $\ell = 1\,$mm, $\rho=20$g/cm$^3$, $m=m_{\rm Cs}$.}
\end{SCfigure*}

We now show how to apply our sensing protocol to a test of quantum gravity. The idea is to realize the qubit in the Hamiltonian \eqref{hamiltonian} as an optical-lattice atom interferometer \cite{xu2019probing} with a hold-time $\tau$ and splitting $\ell$ between the matter wave packets. The majority of the interferometer time sees the atoms trapped in one of two different potential wells created by the lattice. The atom position thus becomes a two-state system with $\sigma_z$-eigenvalues corresponding to the two locations. The mechanical oscillator has a mass $M$ and fundamental frequency $\omega$. Expanding the Newtonian atom-oscillator potential \eqref{VN}, we then have the total Hamiltonian 
\begin{equation}
\label{hamiltonian-final}
H = \omega a^\dag a - g \sigma_z (a + a^\dag).
\end{equation}
Here, $a,a^\dag$ are oscillator operators, so the second term represents the position-position coupling. The coupling strength is
\begin{equation}
g = \kappa \frac{G_N m M \ell x_0}{\hbar R^3}
\end{equation}
where $x_0 = \sqrt{\hbar/2M \omega}$ is the ground state oscillator uncertainty, $R$ parametrizes the distance between the oscillator and atom, and $\kappa$ is a dimensionless number of order one which depends on the specific oscillator mass geometry (see Fig. \ref{fig:setup3}).

\begin{table*}
\footnotesize
\begin{spacing}{1}
\begin{tabular}{| >{\raggedright}p{3.5cm} | >{\raggedright}p{5cm} | p{6cm}<{\raggedright} |}
\hline
{\bf Technical challenge} & {\bf Examples} & {\bf Possible strategies} \\
\hline
\hline
Non-gravitational interactions & Van der Waals, stray fields, scattered laser light & Superconducting shielding, place atoms in waveguide \cite{Xin2018fiber}  \\
\hline
Mean field shift & Parasitic atom-atom interactions leading to inhomogeneous dephasing \cite{Jannin2015Meanfield} & Spin-echo techniques \cite{Laudat} (see also appendix \ref{appendix-spinecho}), fermionic atoms (e.g., Yb-171 or 173) \cite{McAlpine,Niederriter} \\
\hline
Exponential decay of signal & Atomic dephasing & Interleaved differential measurement, e.g. by toggling the mass between near and far positions \cite{Jaffe2017subgrav} \\
\hline
Deviations from harmonicity & Time-dependent oscillator frequency, anharmonic perturbations & Keep effective temperature below nonlinear thresholds; change materials, mounting, or frequency \\
\hline
\end{tabular}
\end{spacing}
\caption{Some systematic effects and other perturbations expected in a realistic implementation.}
\label{table:experiment}
\end{table*}

\begin{table*}
\footnotesize
\begin{spacing}{1}
\begin{tabular}{| >{\raggedright}p{3cm} | >{\raggedright}p{3.6cm} | >{\raggedright}p{3.6cm} |  p{3.8cm}<{\raggedright} |}
\hline
{\bf Loophole or pathology} & {\bf Typical sources} & {\bf Problematic behavior allowed} & {\bf Possible solutions} \\
\hline
\hline
Non-gravitational interactions between atom and oscillator & Casimir/van der Waals interactions & Can generate entanglement (reproduce the full desired signal), can generate extra noise & Vary parameters (masses and distance) to check proper scaling with $V = G_N m_1 m_2/r$ law \\
\hline
Stationarity assumption on bath (and/or experimentalist) violated & Explicit time-dependence introduced by experimentalist (e.g.\ spin-echo protocol); low-frequency noise (e.g., gravity gradients, seismic noise) & Violates assumption of theorem in Sec. \ref{section-theorem}. In principle, could mimic collapse and revival & Adjust theorem to allow for bath relaxation timescale; experimentally verify Markovian nature of oscillator noise \\
\hline
Non-locality & Time of interaction for experiment is much longer than light-crossing time $T_{\rm int} \gg T_{com}$ & Allows for non-local, hidden variable model explaining the entanglement (same as Bell test) & Long baseline version? \\
\hline
\end{tabular}
\end{spacing}
\caption{Some loopholes and pathologies in our proposed test.}
\label{table:loopholes}
\end{table*}

The information sensing protocol requires generation of an initial state $\ket{0} + \ket{1}$. This can be generated, e.g., by a pair of Raman pulses separated by a free evolution time \cite{xu2019probing}, by spin-dependent kicks \cite{Jaffe2018SDK}, by optical lattice techniques \cite{Pagel2020lattice}, or by rapidly splitting a single-well potential to the double-well. Measuring in the $\sigma_z$ basis at the end of the protocol corresponds to closing the atom interferometer and counting the atoms in the two output ports. To implement the ``boosted'' protocol of section \ref{section-boosted}, we can use a number of non-gravitational interactions to generate the initial entanglement. For example, a hyperfine or Rydberg atomic state could be magnetically or optically coupled to the oscillator. Entanglement of this type has been recently demonstrated experimentally \cite{karg2020light,thomas2021entanglement}.

Let us consider how we can obtain a visibility change that is large enough to be measured. In order to observe at least one full cycle of decay and revival, we choose $\omega = 2\pi/\tau$ where $\tau$ is the atom hold-time. In this case, the visibility change is given by
\begin{align}
\begin{split}
\Delta V& = \frac{\pi G_N^2m^2\rho }{3\sqrt{2}\ell \omega^3\hbar }\left(8+\bar n\right) \xrightarrow[k_{\rm B} T/\omega \to \infty]{} \frac{\pi }{3\sqrt{2}}K^2, \\
\Delta V_{\rm b} & = 2^{1/4} G_N m \sqrt{\frac{\pi\rho}{3\ell\omega^3\hbar}(8 +\bar n)}  \xrightarrow[k_{\rm B} T/\omega \to \infty]{} \frac{2^{1/4}}{\sqrt{3}} K
\end{split}
\end{align}
in the unboosted and boosted scheme respectively, where
\begin{align}
\begin{split}
\label{Kvals}
K^2 & = \frac{G_N^2m^2\rho k_B T}{\ell \omega^4\hbar^2} \\
&  \approx 1.04\times 10^{-14} \left(\frac{T}{300\,{\rm K}} \right) \left(\frac{\ell}{1\,{\rm mm}}\right)^{-1}\left(\frac{\tau}{10\,{\rm s}}\right)^4.
\end{split}
\end{align}
Here we took a solid density $\rho = 20~{\rm g/cm}^3,$ cesium atoms $m = m_{\rm Cs} = 133~{\rm amu}$, used the four-sphere configuration (Fig. \ref{fig:setup3}) for definiteness, and maximized the coupling $g$ for a given splitting $\ell$ by choosing a sphere radius of $R_s=\ell/(\sqrt{8})$. Longer atomic interrogation times $\tau$ are preferable. This would require a correspondingly low-frequency oscillator, e.g. a mHz-scale torsional pendulum. While 20\,s have been experimentally realized \cite{xu2019probing}, 100\,s may be a reasonable expectation for the future. Using a small matter-wave splitting $\ell$ is desirable, but subject to mechanical constraints. Choosing, e.g., $\ell = 1\,$mm, $L = 1/\sqrt{2}\,$ mm and $R_s = 0.35\,$mm would leave about $0.15\,$ mm free space between the spheres. For $\tau = 100\,$s and $T=300$\,K we obtain $\Delta V \sim 10^{-10}$; but for the boosted scheme, it will be as large as $\Delta V = 7\times 10^{-6}$ (see Fig. \ref{fig:DeltaV}). At the standard quantum limit, this can be detected with $5-\sigma$ significance by running the experiment with $\sim 5\times 10^{11}\,$ atoms (see appendix \ref{appendix-manyatoms} for details on noise scaling with many atoms). Assuming that the experiment has $10^7$ atoms per run, and each run takes 2 minutes, this will be possible in two months total run time. 

Remarkably, this suggests that the experiment may be feasible in the near future. A number of systematic effects and technical issues will need to be understood. We postpone detailed discussion to future work, but flag some likely issues and ways to handle them in Table \ref{table:experiment}.

\section{Implications, loopholes, and conclusions}

\label{section-conclusions}

Our interactive information sensing protocol is a novel strategy for verification of dynamical entanglement generation. While a standard Bell-type test requires measurements on both parts of a bipartite system, our protocol can verify entanglement generation with only single-body measurements. Crucially, the test verifies the ability of an interaction channel to generate entanglement, without needing to directly verify the entanglement of the final state. However, it is important to note that this test is subject to loopholes. Some are analogous to those in standard Bell tests and others are particular to our proposal. We suggest a few of these in Table \ref{table:loopholes}. 

In our view, the most important loophole stems from our time-translation invariance assumption, which we used to write the atom-oscillator dynamics in Lindblad form \eqref{lindblad}. Non-Markovian time dependence introduced by an experimentalist or Maxwell's demon could, in principle, reproduce the observed collapse and revival dynamics. One way to improve the situation would be to reformulate the theorem to include some level of non-Markovianity, for example a bath relaxation time scale. A more robust option would be to prove experimentally that it is simply the Markovian thermalizing channel acting on the mechanical system. Methods for this include precision quantum thermometry \cite{purdy2017quantum}, which can support the hypothesis of detailed balance. In any case, extending the results here beyond the strictly Markovian assumption will be a crucial next step.

The central technical advances suggested here are the interactive sensing protocol and the use of atoms as a sensor. The key advantage of the periodic collapse-and-revival protocol is that it enables a huge enhancement with a thermal state of the mechanical system; understanding if this can be extended beyond the specific context here would be very interesting. While using trapped atoms is perhaps counter-intuitive since it decreases the strength of the signal (the Newton potential), we emphasize that the extremely long coherence lifetime and ability to generate spatially well-separated superpositions of the atoms lead to similar parametric scaling of the overall signal strength.

We have shown how the interactive sensing protocol can be used to test the ability of the gravitational field to communicate quantum information. If the answer is yes, this would constitute the first direct evidence that the gravitational field itself is a quantum mechanical degree of freedom \cite{carney2019tabletop,belenchia2018quantum,Christodoulou:2018cmk,Marshman:2019sne,Galley:2020qsf}. On the contrary, if the answer is negative, the existence of the graviton is ruled out \cite{carney2019tabletop}. The simple estimates of section \ref{section-experiment} suggest that this experiment is feasible with realistic devices, even in the presence of noise. We will present a more detailed proposal and analysis of systematic effects in a future paper.

\section*{Acknowledgements}

We thank Thomas Guff, Jack Harris, John Kitching, and Jess Riedel for discussions. H.M. has been supported in part by Jet Propulsion Laboratory under grant number 1652036, the Office of Naval Research under grant numbers GRANT12980618 and N00014-20-1-2656, as well as the National Science Foundation under grant number 1708160.

\begin{appendix}

\section{Newtonian entanglement from graviton exchange}
\label{appendix-graviton}

For completeness, we review here some standard arguments about the perturbative quantization of gravity and its relation to entanglement generation via the Newton potential \eqref{VN}. Our goal is to explain the standard logic by which one treats small fluctuations of the metric as a quantum field and uses this to make predictions in non-relativistic systems. We do not mean to say that this derivation somehow proves that this is the correct model of low-energy quantum gravity---on the contrary, determining if this is the correct set of predictions is a central objective of the experiment proposed in this paper.

By far the most common and efficient method to compare a field theoretical description to the non-relativistic setting relevant to these experiments is to do a ``matching'' calculation. For example, one can compute scattering amplitudes in the field theory, compare these to the same amplitude computed in a potential scattering model, and thus obtain the effective non-relativistic potential. Since the scattering states form a complete basis for the Hilbert space (other than bound states), if these two calculations agree for all scattering states, we can conclude that the two descriptions are equivalent quantum-mechanically in the regime in which the calculations match.

\begin{figure}[t]
\includegraphics{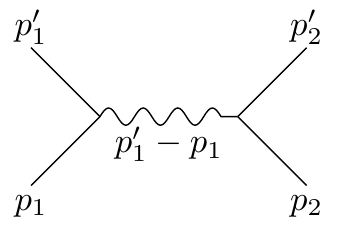}
\includegraphics{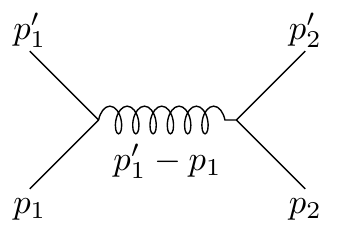}
\caption{Feynman diagrams for single photon and graviton exchange, respectively.}
\label{pic-feynmandiags}
\end{figure}

For example, consider electrodynamics: photons $A_{\mu}$ coupled to a Dirac fermion $\psi$ (e.g., the electron, with charge $e$). See section 4.8 of \cite{peskin2018introduction} for a textbook treatment of what follows. The Lagrangian is
\begin{equation}
\label{Lem}
\mathcal{L} = \frac{1}{4} F_{\mu\nu}F^{\mu\nu} +  \bar{\psi} \gamma^{\mu} (\partial_{\mu} - i e A_{\mu}) \psi
\end{equation}
with $F_{\mu\nu} = \partial_{\mu} A_{\nu} - \partial_{\nu} A_{\mu}$ the electromagnetic field tensor and $\gamma^{\mu}$ the usual Dirac matrices. With this Lagrangian we can perturbatively compute the elastic $e^- e^- \to e^- e^-$ scattering amplitude to lowest order in the charge using standard methods. See the Feynman diagram of figure \ref{pic-feynmandiags}. One finds\footnote{There is also a $u$-channel diagram coming from the indistinguishability of the two electrons, which will also appear in the potential scattering computation, but this is not important for our argument here so we drop it for simplicity.}
\begin{equation}
\braket{\mb{p}_1' \mb{p}_2' | T | \mb{p}_1 \mb{p}_2} = 4\pi e^2 \frac{N_{\rm EM}}{-t - i \epsilon}
\end{equation}
where $t = -(p_1' - p_1)^2$ is the Lorentz-invariant four-momentum transfer, and the numerator is
\begin{equation}
N_{\rm EM} = \bar{u}(\mb{p}_1') \gamma^{\mu} u(\mb{p}_1) \bar{u}(\mb{p}_2') \gamma_{\mu} u(\mb{p}_2).
\end{equation}
Here $u,\bar{u}$ are Dirac spinors. In the non-relativistic limit where the rest masses dominate over the spatial momenta, the numerator reduces to $N_{\rm EM} \to 1$ and $t \to (\mb{p}_1' - \mb{p}_1)^2$. In the center-of-mass frame, this means that the amplitude for the relative momentum $\mb{p} = (\mb{p}_1 - \mb{p}_2)/2$ to transition to $\mb{p}'$ is given by
\begin{equation}
\braket{\mb{p}' | T | \mb{p} } =  \frac{4\pi e^2}{(\mb{p}'-\mb{p})^2},
\end{equation}
where we used conservation of the total momentum to write $\mb{p}_1'-\mb{p}_1 = \mb{p}'-\mb{p}$.

We can then compare this to the amplitude computed non-relativistically with a Hamiltonian description
\begin{equation}
H = H_1 + H_2 + V(r)
\end{equation}
including a central potential $V(r) = V(|\mb{x}_1 - \mb{x}_2|)$. In the center of mass frame, the first Born approximation gives
\begin{equation}
\label{born}
\braket{\mb{p}' | T | \mb{p}} =  \tilde{V}(\mb{p}'-\mb{p}),
\end{equation}
where $\tilde{V}$ is the Fourier transform of the potential. Comparing these expressions, we see that the effective potential is
\begin{equation}
\tilde{V}(\mb{q}) = \frac{4\pi e^2}{\mb{q}^2},
\end{equation}
or in position space,
\begin{equation}
\label{Vofr}
V(\mb{x}) = \int \frac{d^3\mb{q}}{(2\pi)^3} e^{i \mb{q} \cdot \mb{x}} \tilde{V}(\mb{q}) = \frac{e^2}{|\mb{x}|},
\end{equation}
recovering the usual Coulomb force.

We emphasize that this line of argument is at the level of transition \emph{amplitudes} $\braket{\psi_f | U(t_f,t_i) | \psi_i}$, where $U$ means time evolution and we take $t_i \to -\infty, t_f \to +\infty$. Thus, the potential \eqref{Vofr} is the potential operator in the full meaning of the term---it is a two-body operator and can entangle particles, and so forth. Indeed, entanglement generated by this Coulomb potential is exactly what underlies, for example, quantum information processing with chains of trapped ions in Coulomb crystals.

The same type of matching calculation can be performed in gravity. The additional complication is that we need to work perturbatively around a background spacetime, and only quantize the perturbations. For an experiment well-localized in spacetime like ours, we can expand around a locally flat spacetime metric $\eta_{\mu\nu}$, by the equivalence principle.\footnote{A more accurate statement would be that we should expand around a Schwarzschild solution, with Schwarzschild mass given by the mass of Earth. This term would generate a constant background external potential and is inessential to the core quantum mechanics of the following argument.} We write
\begin{equation}
g_{\mu\nu} = \eta_{\mu\nu} + \frac{h_{\mu\nu}}{m_{\rm Pl}}
\end{equation}
where we scale out a factor of the Planck mass to give $h_{\mu\nu}$ dimensions of mass (i.e. the same dimensions as a canonical bosonic field in four dimensions). In this expansion, the Einstein-Hilbert Lagrangian becomes
\begin{equation}
\mathcal{L} = \frac{1}{2} \partial_{\alpha} h_{\mu\nu} \partial^{\alpha} h^{\mu\nu} + \cdots + \frac{1}{m_{\rm Pl}} h_{\mu\nu} T^{\mu\nu},
\end{equation}
where the dots represent other similar second-derivative kinetic terms for the perturbations $h_{\mu\nu}$ and $T_{\mu\nu}$ is the matter stress tensor. These metric perturbations $h_{\mu\nu}$ can be quantized through the exact same procedure as the electromechanical potential $A_{\mu}$ \cite{Feynman:1963ax,t1974one,deser1974one,Veltman:1975vx}. There is a key difference---namely that the interaction is non-renormalizable---which means that we have an effective quantum field theory \cite{kadanoff1966scaling,wilson1971renormalization,weinberg1979phenomenological,donoghue1994general,burgess2004quantum} which can only make reliable predictions at energy densities well below the Planck scale.\footnote{To be more precise: the non-renormalizable nature of the interaction means that we have to include all possible generally-covariant terms, in particular curvature-curvature couplings, in the action $S = m_{\rm Pl}^{2} \int d^4x \sqrt{-g} [ R + m_{\rm Pl}^{-2} (c_1 R^2 + c_2 R_{\mu\nu} R^{\mu\nu} ) + m_{\rm Pl}^{-4} c_3 R^3 + \cdots ]$. Since this is an infinite series of terms each with an unknown constant coefficient $c_i$, the model becomes non-predictive once these $R^{n \geq 2}$ terms become important. In the regime of these experiments $R \ll m_{\rm Pl}^2$, so all of these terms are extraordinarily small and the dynamics is determined entirely by the first term $R$, i.e., the usual Einstein-Hilbert action.}  However, we are well within this limit in the kind of experiment envisaged here, as discussed in the introduction.

Proceeding accordingly, the scattering of a pair of masses (here modelled as single-particle excitations of a massive spinless field) via gravitons is given by \cite{donoghue1994general,burgess2004quantum}
\begin{equation}
\braket{\mb{p}_1' \mb{p}_2' | T | \mb{p}_1 \mb{p}_2}  = \frac{4\pi}{m_{\rm Pl}^2} \frac{N_{\rm grav}}{-t - i \epsilon}.
\end{equation}
The numerator is more complicated due to the tensorial nature of the interaction,
\begin{align}
\begin{split}
N_{\rm grav} & = 2 (p_1 \cdot p_1') (p_2 \cdot p_2') + 2 (p_1 \cdot p_2') (p_1' \cdot p_2) \\
& \ \ \ + 8 (p_1 \cdot p_1' + m^2) (p_2 \cdot p_2' + m^2)
\end{split}
\end{align}
but reduces in the non-relativistic limit to the simple value $N_{\rm grav} \to m^2$. Recognizing that $m_{\rm Pl}^2 = 1/G_N$ in terms of the Newton constant, we can compare this again to the Born approximation \eqref{born}, and determine the effective potential
\begin{equation}
\tilde{V}(\mb{q}) = \frac{4 \pi G_N m^2}{\mb{q}^2}
\end{equation}
which again is just $V(r) = G_N m^2/r$ in real space. In this way, we see that ``graviton exchange'' leads to the Newton potential operator \eqref{VN} in the non-relativistic limit.

Finally, we note that can one directly obtain a Hamiltonian operator for the field theory and read off the non-relativistic potential directly, without resorting to scattering or other matching calculations. In contrast to the gauge-invariant scattering amplitude approach, this is complicated by the gauge symmetries of the model (in both the electrodynamics and gravity cases). To see how this works, consider the electrodynamical Lagrangian \eqref{Lem}. To perform the transformation from the Lagrangian to Hamiltonian we have to fix a gauge, say Coulomb gauge $\partial_i A^i = 0$. This gauge leads to a second-class Dirac constraint $\partial_i F^{i0} = - J^0$, so that $\nabla^2 A^0 = -J^0$, i.e., the $A^0$ part of the potential is non-dynamical and simply fixed by the current
\begin{equation}
A^0(\mb{x},t) = -\int d^3\mb{y} \frac{J^0(\mb{y},t)}{|\mb{x}-\mb{y}|}.
\end{equation}
Performing the Legendre transformation to obtain the Hamiltonian, one then finds a coupling
\begin{align}
\begin{split}
H_{\rm Coul} & = -\int d^3\mb{x} A^0(\mb{x}) J^0(\mb{x}) \\
& = \int d^3\mb{x} d^3\mb{y} \frac{J^0(\mb{x}) J^0(\mb{y})}{|\mb{x}-\mb{y}|},
\end{split}
\end{align}
which is just the usual non-relativistic Coulomb interaction, since $J^0$ is the charge density. Again, everything here is at the level of operators. In the case of gravity, the exactly analogous calculation can be performed and one finds an instantaneous Newton interaction $H_{\rm} = m_{\rm Pl}^{-2} \int d^3\mb{x} d^3\mb{y} T^{00}(\mb{x}) T^{00}(\mb{y})/{|\mb{x}-\mb{y}|}$.  See chapter 8 of \cite{weinberg1995quantum} for a thorough treatment of this kind of Hamiltonian approach in the case of electrodynamics, and e.g. \cite{anastopoulos2013master} for the calculation in perturbative gravity.

The fact that the field component responsible for the Coulomb/Newton interaction is non-dynamical (e.g., $A^0$ in the above example) has led some authors to argue that observing Newtonian entanglement would tell us nothing about the quantization of the ``physical'' (i.e., dynamical) degrees of freedom of the gravitational field; for a prototypical expression of this view, see \cite{Anastopoulos:2018drh}. There are, however, strong arguments against this \cite{belenchia2018quantum,Christodoulou:2018cmk,Marshman:2019sne,Galley:2020qsf}, which essentially say that there is no consistent way to have both an entangling Newton interaction and non-quantized metric fluctuations. We anticipate substantial further debate on this topic, and will present a detailed discussion in a separate paper \cite{carneyprep}.

\section{Detailed calculation of oscillator noise}
\label{appendix-noise}

To get a quantitative estimate of the effects of thermal loading on the oscillator, let us assume that we can safely neglect atomic dephasing $\gamma_a t \ll 1$ for the experimental timescale of interest. To develop an exact result, we will use an input-operator method, in which we include an explicit heat bath for the oscillator. The Lindblad system \eqref{lindblad-noise} can be derived through this method by tracing out the oscillator bath. See \cite{clerk2010introduction} for a lucid review of this technique. 

Let $H_0$ denote the Hamiltonian for the oscillator \emph{and its bath}. The total time evolution operator is then
\begin{equation}
\label{Uopnoise}
U(t) = e^{-i(H_0 + g \sigma_z x)t} = e^{-i H_0 t} \hat{T} e^{-i g \sigma_z \int_0^t x_I(t') dt'}.
\end{equation}
Here $x_I(t)$ is the oscillator position operator in the interaction picture and $\hat{T}$ is the time-ordering operator. For the case of a linear bath, such as that assumed in quantum optics or in quantum Brownian motion, we can explicitly find $x_I$. Writing $x = (a + a^\dag)/\sqrt{2}$, we have
\begin{align}
\begin{split}
\label{aop}
a_I(t) & = \exp[-i(\omega + \gamma_m/2)t] a(0) \\ 
& + \sqrt{\gamma_m} \int_0^t \exp[-i(\omega + \gamma_m/2)(t-t')] a_{\rm in}(t') dt'
\end{split}
\end{align}
where $a_{\rm in}(t)$ is the vacuum noise fluctuation operator, satisfying $[a_{\rm in}(t),a_{\rm in}^\dag(t')] = \delta(t-t')$. Using the linearity of this expression and the Baker-Campbell-Hausdorf relation, we then have that
\begin{align}
\begin{split}
& \hat{T} \exp\left(- i g \sigma_z \int_0^\tau x_I(t) dt\right) \\
& =  \exp\left(- i g \sigma_z \int_0^\tau x_I(t) dt\right) \exp(-i g^2 C(t))
\end{split}
\end{align}
where $C(t)$ is a real, time-dependent number, arising from the non-commuting elements of $x_I(t)$. 

Having dispensed with the time-ordering, we can now explicitly perform the time integral (including a change of integration order in the $a_{\rm in}$ term). Dropping the $e^{-i g^2 C(t)}$ phase, which will cancel out of our observable, we find the time evolution reduces to a simple product of displacement operators, one for the oscillator and one for each mode $a_{\rm in}(t')$ for $0 \leq t' \leq t$, that is
\begin{equation}
U(t) = e^{-i H_0 t} D_a[\sigma_z \alpha(t)] \prod_{0 \leq t' \leq t} D_{a_{in}(t')}[\sigma_z \alpha_{\rm in}(t')]
\end{equation}
where
\begin{align}
\begin{split}
\alpha(t) & = \frac{ig}{i \omega - \gamma_m/2} (1 - e^{(i \omega - \gamma_m/2) t}), \\
\alpha_{\rm in}(t') & = \frac{ig}{i \omega - \gamma_m/2} (1 - e^{(i \omega - \gamma_m/2) (t-t')}).
\end{split}
\end{align}
Finally, we can evaluate our visibility $\sigma_-(t) = U^\dag(t) \sigma_- U(t)$, assuming an initial thermal state for the oscillator and each bath mode and the $\ket{+}$ state for the atom. Using the same results for coherent states as above, one finds
\begin{align}
\begin{split}
\label{analyticnoise}
\braket{\sigma_-} &  = \braket{D_a[ 2 \alpha(t)]}  \prod_{t'} \braket{D_{a_{\rm in}(t')} [\alpha_{\rm in}(t')]} \\
& = \exp\left[- 8\lambda^2 (2 \bar{n} + 1) f(t) \right]
\end{split}
\end{align}
with
\begin{align}
\begin{split}
f(t) & = \frac{\omega^2/4}{\omega^2 + \gamma_m^2/4} \Big( 2 - 2 \cos(\omega t) e^{-\gamma_m t/2} + \gamma_m t \\
& - \frac{8 \gamma_m}{\omega}\sin(\omega t) e^{-\gamma_m t/2} + O(1/Q^2) \Big) 
\end{split}
\end{align}
where $Q = \omega/\gamma_m$ is assumed much larger than one. In particular, at full and half-period this gives
\begin{align}
\begin{split}
V(\pi/\omega) & = \exp [-8 \lambda^2 (2\bar{n}+1)],  \\
V(2 \pi/\omega) & =  \exp [-8 \lambda^2 (2\bar{n}+1)/Q].
\end{split}
\end{align}
Here we have assumed the mechanical damping factor $Q = \omega/\gamma_m \gg 1$. Re-inserting the exponential damping factor for atomic dephasing then reproduces the results in \eqref{SNR}.

\begin{figure}[t]
\includegraphics[width=.85\linewidth]{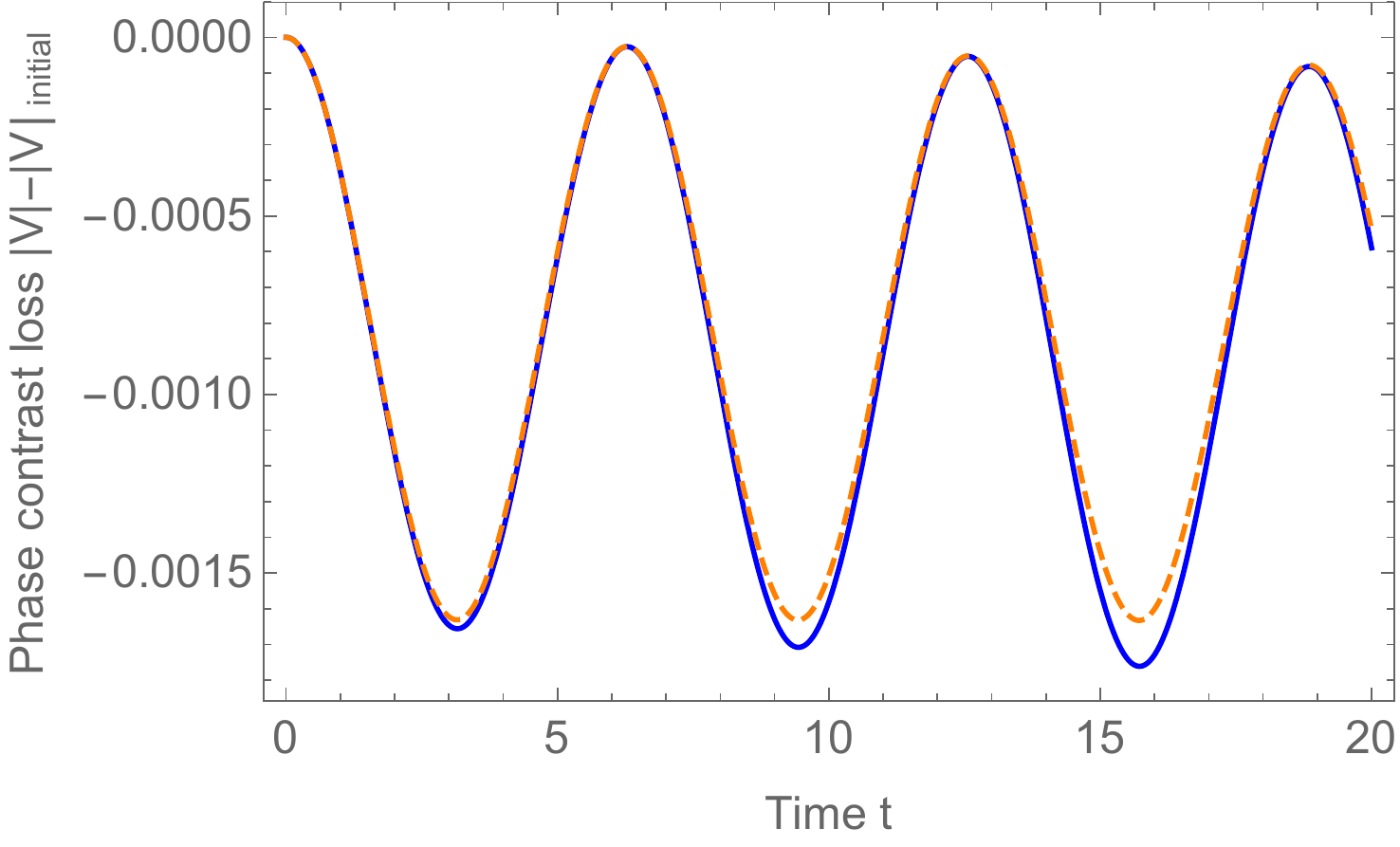} \\ \includegraphics[width=.85\linewidth]{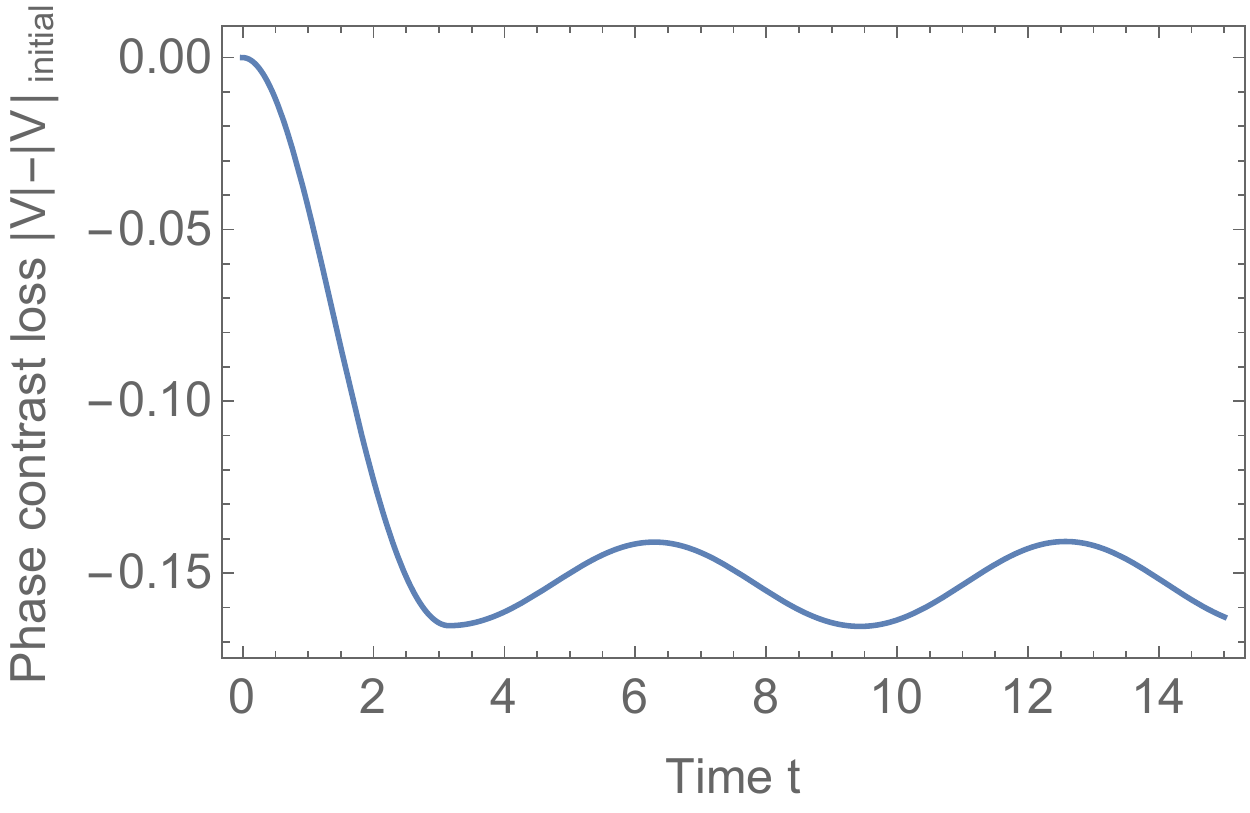}
\caption{Examples of the signal of interest, the phase contrast $V = |\braket{ \sigma_-(t)}|$, compared with its initial value $V(0) = 1/2$. Left: Direct simulation of the Lindblad evolution \eqref{lindblad-noise} in blue, and we see good agreement with our analytic solution including noise \eqref{analyticnoise} in orange, dashed. Normalizing all units to the oscillator frequency $\omega = 1$, here we use values $g = 10^{-2}$ for the gravitational coupling, $\gamma_m = 5 \times 10^{-3}$ for the mechanical damping, and $T = 2$ for the temperature. (Numerical simulation with a much higher $T \gg \omega$ as discussed in the paper is infeasible due to restrictions on the oscillator Hilbert space dimension). Same parameters as left figure, but with an initial $\pi$-pulse using a non-gravitational coupling $g' = 10^{-1}$. The difference between the first collapse and revival is now much larger than in the un-boosted protocol, as predicted in \eqref{SNR-thermalboosted}.}
\label{fig:numerics}
\end{figure}

In Fig. \ref{fig:numerics}, we compare this analytical model with a numerical simulation of the Lindblad equation \eqref{lindblad}, showing excellent agreement.

\section{Detailed calculation of the boosted protocol}
\label{appendix-boostedcalc}

Here we give the full computation of the visibility in our entanglement-enhanced, ``boosted'' protocol of Section \ref{section-boosted}. The total evolution is a product of two unitaries, one for the first half-period under the coupling $g+g'$ and the second under only $g$. We write these as
\begin{align}
\begin{split}
U_{g+g'} & = D^\dag((\lambda+\lambda')\sigma_z) e^{-i \omega n} D((\lambda+\lambda')\sigma_z), \\
U_{g}(t) & =  D^\dag(\lambda\sigma_z) e^{-i \omega n(t-\pi/\omega)} D(\lambda \sigma_z).
\end{split}
\end{align}
With this notation, the visibility of the atom, given some initial coherent state $\ket{\alpha}$ for the oscillator, is given by (defining $\tilde{\lambda} = \lambda + \lambda'$ for brevity)
\begin{align}
\begin{split}
\label{Vlongcalc}
V_{{\rm b},\alpha}(t) & = \braket{\alpha | U^\dag_{g+g'} U_{g}(t) \sigma_- U_{g}(t) U_{g+g'}(t) | \alpha} \\
& = \bra{\alpha} D^\dag(-\tilde{\lambda}) e^{i \omega n/\pi} D(-\tilde{\lambda}) U^\dag_g(t) \\
& \ \ \ \times \sigma_- U_g(t) D^\dag(\tilde{\lambda}) e^{-i \omega n/\pi} D(\tilde{\lambda}) \ket{\alpha} \\
& = \braket{\alpha | D(2\tilde{\lambda}) e^{i \omega n/\pi}  U^\dag_g(t) \sigma_- U_g(t) e^{-i \omega n/\pi} D(2\tilde{\lambda}) | \alpha} \\
& = \braket{\alpha | D\left[ 2\tilde{\lambda} - \lambda(1+e^{i \omega t}) \right] D\left[ 2\tilde{\lambda} - \lambda(1+e^{i \omega t}) \right] | \alpha} \\
& = \bra{0} D(-\alpha) D\left[ 2\tilde{\lambda} - \lambda(1+e^{i \omega t}) \right] \\
& \ \ \ \times D\left[ 2\tilde{\lambda} - \lambda(1+e^{i \omega t}) \right] D(\alpha) \ket{0} \\
& = e^{\phi} \braket{ \alpha - 2\tilde{\lambda} + \lambda(1+e^{i \omega t}) | \alpha + 2\tilde{\lambda} - \lambda(1+e^{i \omega t}) } \\
& = e^{2 \phi} e^{-|4 \tilde{\lambda}^2 - 2\lambda(1+e^{i \omega t})|^2/2}. 
\end{split}
\end{align}
To go from the second to third line, we inserted a pair of identity operators $1 = e^{-i \omega n/\pi} e^{i \omega n/\pi}$ and used $e^{i \omega n/\pi} D(\tilde{\lambda}) e^{-i \omega n/\pi} = D(-\tilde{\lambda})$. From the third to fourth we used the same trick and the more general time evolution $e^{i \omega n t} D(\tilde{\lambda}) e^{-i \omega n t} = D(\tilde{\lambda}e^{i \omega t})$. In the last few lines the ``phase'' is
\begin{equation}
\phi = \alpha^* (2 \tilde{\lambda} -\lambda(1+e^{i \omega t}))/2 + {\rm c.c.}.
\end{equation}
Note that we got two factors of this: one in the fifth line, from the braiding relation $D(\alpha) D(\beta) = e^{(\alpha \beta^* - \alpha^* \beta)/2}$, and then another in the subsequent line from the inner product $\braket{\beta| \alpha} = e^{-|\beta-\alpha|^2/2} e^{(\alpha \beta^* - \alpha^* \beta)/2}$. Notice also that the second exponential does not depend on the coherent state parameter $\alpha$. Thus we only need to average this phase term over the Glauber representation, which gives
\begin{align}
\begin{split}
& \int \frac{d^2\alpha}{\pi \bar{n}} e^{-|\alpha|^2/\bar{n}} e^{2 \phi} \\
& \ \ \ = \exp \left(8 \lambda  n (\lambda +2 \lambda' ) \cos (t \omega )-8 n \left(\lambda ^2+2 \lambda  \lambda' +2 \lambda'^2\right)\right),
\end{split}
\end{align}
where we used the explicit coefficient $\tilde{\lambda} = \lambda + \lambda'$. Doing the same with the second term in \eqref{Vlongcalc}, and simplifying the terms, we finally obtain
\begin{align}
\begin{split}
& V_{\rm b}(t) = \int \frac{d^2\alpha}{\pi \bar{n}} V_{{\rm b},\alpha}(t) \\ 
& = \exp\left[ -8(2\bar{n}+1) \left(  \lambda'^2 + 2 \lambda \lambda' \sin^2 \frac{\omega t}{2} + \lambda^2 \sin^2 \frac{\omega t}{2} \right) \right],
\end{split}
\end{align}
as quoted in \eqref{signal-boosted-timeseries}. Note the limit $\lambda' \to 0$ reproduces the basic, un-boosted protocol. We show the form of this visibility evolution in Fig. \ref{fig:numerics}.

\section{Using many atoms}
\label{appendix-manyatoms}

\begin{figure*}[ht]
\centering
\includegraphics[width=.95\linewidth]{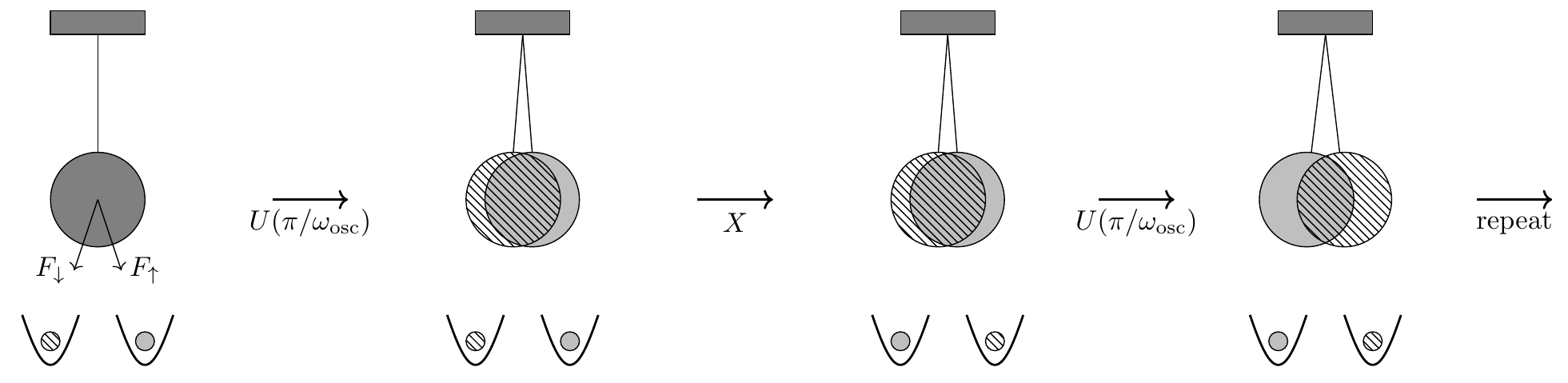} \vspace{.2in} \\
\caption{Spin-echo variant of the basic protocol. After a half-period of evolution, the two pathways through oscillator phase space are maximally distant. The atomic positions are then flipped ($X$ gate), followed by another half-period of evolution. This procedure can be repeated arbitrarily, leading to a net amplification of the basic protocol. The bottom-right figure shows the resulting conditioned paths of the oscillator through phase space, through one iteration.}
\label{fig:spin-echo}
\end{figure*}

The sensitivity of the protocol can be substantially improved by moving from using a single atom to use a collection of atoms, as is typical in an atom interferometer \cite{kasevich1992measurement,santarelli1999quantum,gross2010nonlinear}. For simplicity, we will take $g$ the same for all the atoms, though that is not necessary in practice. In that limit, we can define $J_z = \sum_j \sigma_z^j$ and $J_- = \sum_j \sigma_-^j$ as the collective variables that will enter.

Consider the extension of \eqref{visibility} to the case of $N$ atoms prepared in the initial state $\ket{+++\cdots}$. The ``observable'' of interest is $\braket{J_-(t)}$. This is easiest to calculate term by term for each atom. The total time evolution operator, following the same logic as in \eqref{Uop}, is
\begin{equation}
\label{Uopappendix}
U(t) = \exp \left( -i \beta J_z^2 \right) D^\dag(\lambda J_z) e^{i \omega a^\dag a t} D(\lambda J_z),
\end{equation}
with
\begin{equation}
\beta = \frac{g^2 t}{\omega}.
\end{equation}
This total $J_z^2$ term is peculiar to the case of $N > 1$ atoms; for $N = 1$ it is just an overall phase which we dropped in \eqref{Uop}. Here, however, it is a non-trivial operator, physically representing the ponderomotive squeezing of the spins due to the gravitational coupling with the oscillator. At time $t > 0$ we have, for each $i = 1,\ldots,N$, the operator evolution
\begin{equation}
\label{sigma-evo}
\sigma_-^i(t) = U^\dag(t) \sigma_-^i U(t).
\end{equation}
If we define $\tilde{J}_z^i = J_z - \sigma_z^i$, we can see that we can write the $J_z^2$ contribution to \eqref{sigma-evo} as
\begin{equation}
\label{phasecommutator}
\exp \left( i \beta J_z^2 \right) \sigma_-^i \exp \left( -i \beta J_z^2 \right) = \exp \left( - 2 i \beta \tilde{J}_z \right) \sigma_-^i.
\end{equation}
This prefactor then commutes with the rest of the operators in \eqref{sigma-evo}. Using this and the same basic logic as in \eqref{visibility}, we find that all the $i \neq j$ spins will just give a phase proportional to $\tilde{J}_z$:
\begin{equation}
\sigma_-^i(t) = \sigma_-^i e^{-2 i \beta \tilde{J}_z} D(-\lambda) D(2 \lambda e^{i \omega t}) D(-\lambda).
\end{equation}
Acting on the initial state $\ket{0,+ + + \cdots}$ with the oscillator prepared in $\ket{0}$ and each atom in the $\ket{+}$ state, that is to say with $N$ unentangled atoms, we obtain
\begin{equation}
\label{sigma-final}
\braket{\sigma_-^i(t)} = \cos^{N-1} ( 2 \beta) \braket{ \sigma_-(t)}_1
\end{equation}
where the term $\braket{\sigma_-(t)}_1$ means the answer with a single spin, as in equations \eqref{visibility} and \eqref{sigma-}. For $N \gg 1$ and $\beta \ll 1$ (recall $\beta = g^2 t/\omega$, so this condition is certainly satisfied for us), we can Taylor expand the cosine and match it to an exponential for convenience, i.e., $\cos^{N-1} (2 \beta) \approx e^{- 2 N \beta^2}$. Thus since $\braket{J_-}$ has $N$ terms of the form \eqref{sigma-final}, we finally get
\begin{equation}
\braket{J_-(t)} = N e^{-2 N g^4 t^2/\omega^2} \braket{\sigma_-(t)}_1.
\end{equation}
For example, at a half period and full period we then have
\begin{align}
\begin{split}
\label{result-phasenoise}
\braket{J_-(\pi/\omega)} & = N e^{-2 \pi^2 N \lambda^4} e^{-8 \lambda^2}, \\
\braket{J_-(2 \pi/\omega)} & = N e^{-4 \pi^2 N \lambda^4}.
\end{split}
\end{align}
Note that this noisy phase is independent of the oscillator's initial state, so for example we get the same answer if the oscillator begins in a thermal state. We see the basic $N$ enhancement to the signal here as the prefactor. The phase noise scales like $N \lambda^4$. For our particular implementation with parameters like those quoted in \eqref{Kvals}, we have $\lambda \sim 10^{-13}$, so for $N \sim 10^{10}$ atoms these phase noise exponentials are completely negligible. The overall $N$ factor here represents the usual $\sqrt{N}$ statistical enhancement in the signal-to-noise, assuming uncorrelated atom errors.

The same calculation extends directly to the entanglement-enhanced, $g$-linear protocol \eqref{signal-boosted-timeseries}. This is clear by the algebraic structure of the argument given above. Explicitly, we now have two time evolution operators of the form \eqref{Uopappendix}, one with a coupling $g+g'$ from $t=-\pi/\omega$ to $t=0$, followed by another with only the gravitational $g$ coupling from $t=0$ onward. In an obvious notation we can write
\begin{equation}
\sigma^i_-(t) = U^\dag_{g}(t) U^\dag_{g+g'} \sigma_-^i U_{g+g'} U_{g}(t).
\end{equation}
In these $U$ operators, we have the same phase-noise terms, namely $e^{i \beta J_z^2}$ in the $U_g$ and another factor $e^{i \beta' J_z^2}$, with $\beta' = -(g+g')^2 \pi/\omega$, from the $U_{g+g'}$ factor. These depend only on the $J_z$ operator and thus commute with the other terms (displacement operators and free oscillator evolution) in $U_g$ and $U_{g+g'}$. Thus we get an expression
\begin{align}
\begin{split}
\sigma^i_-(t) & \sim \exp( i (\beta+\beta') J_z^2 ) \sigma_-^i \exp( -i (\beta+\beta') J_z^2) \\
& = \exp(-2 i (\beta+\beta') \tilde{J}_z) \sigma_-^i,
\end{split}
\end{align}
just as in \eqref{phasecommutator} except now with $\beta$ replaced by
\begin{equation}
\beta + \beta' = \frac{g^2 t}{\omega} - \frac{(g+g')^2 \pi}{\omega}.
\end{equation}
In particular, all the terms other than these phase noise exponentials contribute as given by \eqref{signal-boosted-timeseries}. The overall signal is still increased linearly in $N$ as in \eqref{result-phasenoise}, times a negligible contribution from this ponderomotive squeezing noise.

\section{Spin-echo version for faster physical oscillators}
\label{appendix-spinecho}

The physical oscillator frequency is crucially important to the size of the observable effect. The interferometric contrast scales as a power of $\lambda = g/\omega$, so a low-frequency oscillator is ideal. However, in practice, using a very low-frequency (sub-Hz) oscillator would present substantial technical problems. This can be alleviated by using a high-frequency oscillator and a spin-echo like sequence to mimic the effect of a low-frequency oscillator. Specifically, after every $\pi/\omega$ half-period, we swap the two atom locations, i.e.\ perform a $\sigma_x$ operation (see Fig. \ref{fig:spin-echo}). This produces the evolution
\begin{align}
\begin{split}
U &= \sigma_x e^{-i H \tau/2} \sigma_x e^{-i H \tau/2} \\
&= e^{-i (H_p - V) \tau/2} e^{-i (H_p + V) \tau/2} \\
&= D(\lambda \sigma_z) e^{-i H_p \tau/2} D(-\lambda \sigma_z) D(-\lambda \sigma_z) e^{-i H_p \tau/2} D(\lambda \sigma_z) \\
&= D(\lambda \sigma_z) D(2 \lambda \sigma_z) D(\lambda \sigma_z) = D(4 \lambda \sigma_z)
\end{split}
\end{align}
where we used $\tau = 2 \pi/\omega$ for the final line. With $N_{\pi}$ iterations of this, we produce the total evolution $U(N_{\pi}) = D(4 N_{\pi} \lambda \sigma_z)$. Performing $N_{\pi}$ iterations, followed by a $\sigma_x$ operation, followed by $N_{\pi}$ further iterations, we recover the revival:
\begin{align}
\begin{split}
U_{\rm spin-echo} & = \sigma_x D(4 N_\pi \lambda \sigma_z) \sigma_x D(4 N_\pi \lambda \sigma_z) \\
& = D(-4 N_\pi \lambda \sigma_z) D(4 N_\pi \lambda \sigma_z) \\
& = \mathbb{I}.
\end{split}
\end{align}
In this spin-echo style variant, the wavefunction overlap after a total time $t = N_{\pi} \tau$, i.e.\ to the halfway point, is given by 
\begin{equation}
O = \exp\left[-32 \frac{N_{\pi}^2 g^2}{\omega^2}
\right].
\end{equation}
Thus we have an effect scaling like $g^2/\omega_{\rm eff}^2$, with the effective frequency $\omega_{\rm eff} = 2\pi/t = \omega/N_{\pi}$. This means we can have an effectively slow oscillator (which is beneficial for the signal strength) while using a faster physical oscillator (beneficial for noise reasons), at the cost of having to perform some $\sigma_x$ operations on the two-state system.

We note that application of this spin-echo protocol would violate the Markov assumption used to prove the theorem in section \ref{section-theorem}. Adjusting that proof to accommodate this specific type of non-Markovian control would be necessary to draw the same conclusion, namely, that the gravitational entanglement in this protocol is necessarily due to the mass-atom entangling interaction.

\end{appendix}

\bibliographystyle{utphys-dan}
\bibliography{refs}

\providecommand{\href}[2]{#2}\begingroup\raggedright\begin{thebibliography}{10}

\bibitem{feynman1971lectures}
R.~P. Feynman, F.~B. Morinigo, and W.~G. Wagner, {\em Lectures on Gravitation,
  1962-63}.
\newblock California Institute of Technology, 1971.

\bibitem{page1981indirect}
D.~N. Page and C.~Geilker, ``Indirect evidence for quantum gravity,'' {\em
  Physical Review Letters} {\bfseries 47} no.~14, (1981) 979.

\bibitem{diosi1984gravitation}
L.~Di{\'o}si, ``Gravitation and quantum-mechanical localization of
  macro-objects,'' {\em Physics Letters, Section A: General, Atomic and Solid
  State Physics} {\bfseries 105} no.~4-5, (1984) 199--202.

\bibitem{penrose1996gravity}
R.~Penrose, ``On gravity's role in quantum state reduction,'' {\em General
  relativity and gravitation} {\bfseries 28} no.~5, (1996) 581--600.

\bibitem{kafri2015bounds}
D.~Kafri, G.~Milburn, and J.~Taylor, ``Bounds on quantum communication via
  {N}ewtonian gravity,'' {\em New Journal of Physics} {\bfseries 17} no.~1,
  (2015) 015006.

\bibitem{bose2017spin}
S.~Bose, A.~Mazumdar, G.~W. Morley, H.~Ulbricht, M.~Toro{\v{s}},
  M.~Paternostro, A.~A. Geraci, P.~F. Barker, M.~Kim, and G.~Milburn, ``Spin
  entanglement witness for quantum gravity,'' {\em Physical Review Letters}
  {\bfseries 119} no.~24, (2017) 240401.

\bibitem{Marletto:2017kzi}
C.~Marletto and V.~Vedral, ``{Gravitationally-induced entanglement between two
  massive particles is sufficient evidence of quantum effects in gravity},''
  \href{http://dx.doi.org/10.1103/PhysRevLett.119.240402}{{\em Phys. Rev.
  Lett.} {\bfseries 119} no.~24, (2017) 240402},
  \href{http://arxiv.org/abs/1707.06036}{{\ttfamily arXiv:1707.06036
  [quant-ph]}}.

\bibitem{Haine:2018bwu}
S.~A. Haine, ``{Searching for Signatures of Quantum Gravity in Quantum
  Gases},'' \href{http://arxiv.org/abs/1810.10202}{{\ttfamily arXiv:1810.10202
  [quant-ph]}}.

\bibitem{Chevalier:2020uvv}
H.~Chevalier, A.~J. Paige, and M.~S. Kim, ``{Witnessing the nonclassical nature
  of gravity in the presence of unknown interactions},''
  \href{http://dx.doi.org/10.1103/PhysRevA.102.022428}{{\em Phys. Rev. A}
  {\bfseries 102} no.~2, (2020) 022428},
  \href{http://arxiv.org/abs/2005.13922}{{\ttfamily arXiv:2005.13922
  [quant-ph]}}.

\bibitem{Howl:2020isj}
R.~Howl, V.~Vedral, M.~Christodoulou, C.~Rovelli, D.~Naik, and A.~Iyer,
  ``{Testing Quantum Gravity with a Single Quantum System},''
  \href{http://arxiv.org/abs/2004.01189}{{\ttfamily arXiv:2004.01189
  [quant-ph]}}.

\bibitem{Anastopoulos:2020cdp}
C.~Anastopoulos and B.-L. Hu, ``{Quantum Superposition of Two Gravitational Cat
  States},'' \href{http://dx.doi.org/10.1088/1361-6382/abbe6f}{{\em Class.
  Quant. Grav.} {\bfseries 37} no.~23, (2020) 235012},
  \href{http://arxiv.org/abs/2007.06446}{{\ttfamily arXiv:2007.06446
  [quant-ph]}}.

\bibitem{Matsumura:2020law}
A.~Matsumura and K.~Yamamoto, ``{Gravity-induced entanglement in optomechanical
  systems},'' \href{http://dx.doi.org/10.1103/PhysRevD.102.106021}{{\em Phys.
  Rev. D} {\bfseries 102} no.~10, (2020) 106021},
  \href{http://arxiv.org/abs/2010.05161}{{\ttfamily arXiv:2010.05161 [gr-qc]}}.

\bibitem{carney2019tabletop}
D.~Carney, P.~C.~E. Stamp, and J.~M. Taylor, ``Tabletop experiments for quantum
  gravity: a user’s manual,'' {\em Classical and Quantum Gravity} {\bfseries
  36} no.~3, (2019) 034001.

\bibitem{HORODECKI19961}
M.~Horodecki, P.~Horodecki, and R.~Horodecki, ``Separability of mixed states:
  necessary and sufficient conditions,'' {\em Physics Letters A} {\bfseries
  223} no.~1, (1996) 1--8.

\bibitem{terhal2000bell}
B.~M. Terhal, ``Bell inequalities and the separability criterion,'' {\em
  Physics Letters A} {\bfseries 271} no.~5-6, (2000) 319--326.

\bibitem{feynman1963theory}
R.~P. Feynman and F.~L. Vernon, ``The theory of a general quantum system
  interacting with a linear dissipative system,'' {\em Annals of Physics}
  {\bfseries 24} no.~11, (1963) 173.

\bibitem{caldeira1983path}
A.~O. Caldeira and A.~J. Leggett, ``Path integral approach to quantum brownian
  motion,'' {\em Physica A: Statistical mechanics and its Applications}
  {\bfseries 121} no.~3, (1983) 587--616.

\bibitem{joos1985emergence}
E.~Joos and H.~D. Zeh, ``The emergence of classical properties through
  interaction with the environment,'' {\em Zeitschrift f{\"u}r Physik B
  Condensed Matter} {\bfseries 59} no.~2, (1985) 223--243.

\bibitem{zurek2003decoherence}
W.~H. Zurek, ``Decoherence, einselection, and the quantum origins of the
  classical,'' {\em Reviews of Modern Physics} {\bfseries 75} no.~3, (2003)
  715.

\bibitem{kasevich1992measurement}
M.~Kasevich and S.~Chu, ``Measurement of the gravitational acceleration of an
  atom with a light-pulse atom interferometer,'' {\em Applied Physics B}
  {\bfseries 54} no.~5, (1992) 321--332.

\bibitem{santarelli1999quantum}
G.~Santarelli, P.~Laurent, P.~Lemonde, A.~Clairon, A.~G. Mann, S.~Chang, A.~N.
  Luiten, and C.~Salomon, ``Quantum projection noise in an atomic fountain: A
  high stability cesium frequency standard,'' {\em Physical Review Letters}
  {\bfseries 82} no.~23, (1999) 4619.

\bibitem{gross2010nonlinear}
C.~Gross, T.~Zibold, E.~Nicklas, J.~Esteve, and M.~K. Oberthaler, ``Nonlinear
  atom interferometer surpasses classical precision limit,'' {\em Nature}
  {\bfseries 464} no.~7292, (2010) 1165--1169.

\bibitem{xu2019probing}
V.~Xu, M.~Jaffe, C.~D. Panda, S.~L. Kristensen, L.~W. Clark, and H.~M{\"u}ller,
  ``Probing gravity by holding atoms for 20 seconds,'' {\em Science} {\bfseries
  366} no.~6466, (2019) 745--749.

\bibitem{lee2020new}
J.~Lee, E.~Adelberger, T.~Cook, S.~Fleischer, and B.~Heckel, ``New test of the
  gravitational $1/r^2$ law at separations down to 52 $\mu$m,'' {\em Physical
  Review Letters} {\bfseries 124} no.~10, (2020) 101101.

\bibitem{catano2020high}
S.~B. Cata{\~n}o-Lopez, J.~G. Santiago-Condori, K.~Edamatsu, and N.~Matsumoto,
  ``High-{Q} milligram-scale monolithic pendulum for quantum-limited gravity
  measurements,'' {\em Physical Review Letters} {\bfseries 124} no.~22, (2020)
  221102.

\bibitem{belenchia2018quantum}
A.~Belenchia, R.~M. Wald, F.~Giacomini, E.~Castro-Ruiz, {\v{C}}.~Brukner, and
  M.~Aspelmeyer, ``Quantum superposition of massive objects and the
  quantization of gravity,'' {\em Physical Review D} {\bfseries 98} no.~12,
  (2018) 126009.

\bibitem{Christodoulou:2018cmk}
M.~Christodoulou and C.~Rovelli, ``{On the possibility of laboratory evidence
  for quantum superposition of geometries},''
  \href{http://dx.doi.org/10.1016/j.physletb.2019.03.015}{{\em Phys. Lett. B}
  {\bfseries 792} (2019) 64--68},
  \href{http://arxiv.org/abs/1808.05842}{{\ttfamily arXiv:1808.05842 [gr-qc]}}.

\bibitem{Marshman:2019sne}
R.~J. Marshman, A.~Mazumdar, and S.~Bose, ``{Locality and entanglement in
  table-top testing of the quantum nature of linearized gravity},''
  \href{http://dx.doi.org/10.1103/PhysRevA.101.052110}{{\em Phys. Rev. A}
  {\bfseries 101} no.~5, (2020) 052110},
  \href{http://arxiv.org/abs/1907.01568}{{\ttfamily arXiv:1907.01568
  [quant-ph]}}.

\bibitem{Galley:2020qsf}
T.~D. Galley, F.~Giacomini, and J.~H. Selby, ``{A no-go theorem on the nature
  of the gravitational field beyond quantum theory},''
  \href{http://arxiv.org/abs/2012.01441}{{\ttfamily arXiv:2012.01441
  [quant-ph]}}.

\bibitem{Feynman:1963ax}
R.~P. Feynman, ``{Quantum theory of gravitation},'' {\em Acta Phys. Polon.}
  {\bfseries 24} (1963) 697--722.

\bibitem{t1974one}
G.~t~Hooft and M.~Veltman, ``One-loop divergencies in the theory of
  gravitation,'' in {\em Annales de l'IHP Physique Th{\'e}orique}, vol.~20,
  pp.~69--94.
\newblock 1974.

\bibitem{deser1974one}
S.~Deser and P.~van Nieuwenhuizen, ``One-loop divergences of quantized
  {E}instein-{M}axwell fields,'' {\em Physical Review D} {\bfseries 10} no.~2,
  (1974) 401.

\bibitem{Veltman:1975vx}
M.~J.~G. Veltman, ``{Quantum Theory of Gravitation},'' {\em Conf. Proc. C}
  {\bfseries 7507281} (1975) 265--327.

\bibitem{donoghue1994general}
J.~F. Donoghue, ``General relativity as an effective field theory: The leading
  quantum corrections,'' {\em Physical Review D} {\bfseries 50} no.~6, (1994)
  3874.

\bibitem{burgess2004quantum}
C.~P. Burgess, ``Quantum gravity in everyday life: General relativity as an
  effective field theory,'' {\em Living Reviews in Relativity} {\bfseries 7}
  no.~1, (2004) 1--56.

\bibitem{howl2019exploring}
R.~Howl, R.~Penrose, and I.~Fuentes, ``Exploring the unification of quantum
  theory and general relativity with a bose--einstein condensate,'' {\em New
  Journal of Physics} {\bfseries 21} no.~4, (2019) 043047.

\bibitem{Tilloy:2019hxe}
A.~Tilloy, ``{Does gravity have to be quantized? Lessons from non-relativistic
  toy models},'' \href{http://dx.doi.org/10.1088/1742-6596/1275/1/012006}{{\em
  J. Phys. Conf. Ser.} {\bfseries 1275} no.~1, (2019) 012006},
  \href{http://arxiv.org/abs/1903.01823}{{\ttfamily arXiv:1903.01823
  [quant-ph]}}.

\bibitem{Bruschi:2020xbm}
D.~E. Bruschi and F.~K. Wilhelm, ``{Self gravity affects quantum states},''
  \href{http://arxiv.org/abs/2006.11768}{{\ttfamily arXiv:2006.11768
  [quant-ph]}}.

\bibitem{Kibble:1979jn}
T.~W.~B. Kibble and S.~Randjbar-Daemi, ``{Nonlinear Coupling of Quantum Theory
  and Classical Gravity},''
\href{http://dx.doi.org/10.1088/0305-4470/13/1/015}{{\em J. Phys.} {\bfseries
  A13} (1980) 141}.

\bibitem{kafri2014classical}
D.~Kafri, J.~Taylor, and G.~Milburn, ``A classical channel model for
  gravitational decoherence,'' {\em New Journal of Physics} {\bfseries 16}
  no.~6, (2014) 065020.

\bibitem{Oppenheim:2018igd}
J.~Oppenheim, ``{A post-quantum theory of classical gravity?},''
  \href{http://arxiv.org/abs/1811.03116}{{\ttfamily arXiv:1811.03116
  [hep-th]}}.

\bibitem{Kent:2020gov}
A.~Kent, ``{Tests of Quantum Gravity near Measurement Events},''
  \href{http://arxiv.org/abs/2010.11811}{{\ttfamily arXiv:2010.11811 [gr-qc]}}.

\bibitem{guhne2009entanglement}
O.~G{\"u}hne and G.~T{\'o}th, ``Entanglement detection,'' {\em Physics Reports}
  {\bfseries 474} no.~1-6, (2009) 1--75.

\bibitem{pezze2018quantum}
L.~Pezze, A.~Smerzi, M.~K. Oberthaler, R.~Schmied, and P.~Treutlein, ``Quantum
  metrology with nonclassical states of atomic ensembles,'' {\em Reviews of
  Modern Physics} {\bfseries 90} no.~3, (2018) 035005.

\bibitem{stern1990phase}
A.~Stern, Y.~Aharonov, and Y.~Imry, ``Phase uncertainty and loss of
  interference: A general picture,'' {\em Physical Review A} {\bfseries 41}
  no.~7, (1990) 3436.

\bibitem{rowan1965electron}
L.~Rowan, E.~Hahn, and W.~Mims, ``Electron-spin-echo envelope modulation,''
  {\em Physical Review} {\bfseries 137} no.~1A, (1965) A61.

\bibitem{dikanov1992electron}
S.~A. Dikanov and Y.~Tsvetkov, {\em Electron spin echo envelope modulation
  (ESEEM) spectroscopy}.
\newblock CRC press, 1992.

\bibitem{raimond2001manipulating}
J.-M. Raimond, M.~Brune, and S.~Haroche, ``Manipulating quantum entanglement
  with atoms and photons in a cavity,'' {\em Reviews of Modern Physics}
  {\bfseries 73} no.~3, (2001) 565.

\bibitem{rains1999rigorous}
E.~M. Rains, ``Rigorous treatment of distillable entanglement,'' {\em Physical
  Review A} {\bfseries 60} no.~1, (1999) 173.

\bibitem{lindblad1976generators}
G.~Lindblad, ``On the generators of quantum dynamical semigroups,'' {\em
  Communications in Mathematical Physics} {\bfseries 48} no.~2, (1976)
  119--130.

\bibitem{gorini1976completely}
V.~Gorini, A.~Kossakowski, and E.~C.~G. Sudarshan, ``Completely positive
  dynamical semigroups of n-level systems,'' {\em Journal of Mathematical
  Physics} {\bfseries 17} no.~5, (1976) 821--825.

\bibitem{preskill1998lecture}
J.~Preskill, ``Lecture notes on quantum information and computation,'' 1998.

\bibitem{gallis1990environmental}
M.~R. Gallis and G.~N. Fleming, ``Environmental and spontaneous localization,''
  {\em Physical Review A} {\bfseries 42} no.~1, (1990) 38.

\bibitem{karg2020light}
T.~M. Karg, B.~Gouraud, C.~T. Ngai, G.-L. Schmid, K.~Hammerer, and
  P.~Treutlein, ``Light-mediated strong coupling between a mechanical
  oscillator and atomic spins 1 meter apart,'' {\em Science} {\bfseries 369}
  no.~6500, (2020) 174--179.

\bibitem{thomas2021entanglement}
R.~A. Thomas, M.~Parniak, C.~{\O}stfeldt, C.~B. M{\o}ller, C.~B{\ae}rentsen,
  Y.~Tsaturyan, A.~Schliesser, J.~Appel, E.~Zeuthen, and E.~S. Polzik,
  ``Entanglement between distant macroscopic mechanical and spin systems,''
  {\em Nature Physics} {\bfseries 17} no.~2, (2021) 228--233.

\bibitem{Xin2018fiber}
M.~X. Xin, W.~S. Leong, C.~Zilong, and S.-Y. Lan, ``An atom interferometer
  inside a hollow-core photonic crystal fiber,'' {\em Science Advances}
  {\bfseries 4} (2018) e1701723.

\bibitem{Jannin2015Meanfield}
R.~Jannin, P.~Clad\'e, and S.~Guellati-Kh\'elifa, ``Phase shift due to
  atom-atom interactions in a light-pulse atom interferometer,'' {\em Physical
  Review A} {\bfseries 92} (2015) 013616.

\bibitem{Laudat}
T.~Laudat, V.~Dugrain, T.~Mazzoni, M.-Z. Huang, C.~L.~G. Alzar, A.~Sinatra,
  P.~Rosenbusch, and J.~Reichel, ``Spontaneous spin squeezing in a rubidium
  bec,'' {\em New Journal of Physics} {\bfseries 20} (2018) 073018.

\bibitem{McAlpine}
K.~E. McAlpine, D.~Gochnauer, and S.~Gupta, ``Excited-band {B}loch oscillations
  for precision atom interferometry,'' {\em Physical Review A} {\bfseries 101}
  (2020) 023614.

\bibitem{Niederriter}
R.~D. Niederriter, C.~Schlupf, and P.~Hamilton, ``Cavity probe for real-time
  detection of atom dynamics in an optical lattice,'' {\em Physical Review A}
  {\bfseries 102} (2020) 051301(R).

\bibitem{Jaffe2017subgrav}
M.~Jaffe, P.~Haslinger, V.~X. Xu, P.~Hamilton, A.~Upadhye, B.~Elder, J.~Khoury,
  and H.~M\"uller, ``Testing sub-gravitational forces on atoms from a
  miniature, in-vacuum source mass,'' {\em Nature Physics} {\bfseries 13}
  (2017) 938--942.

\bibitem{Jaffe2018SDK}
M.~Jaffe, V.~Xu, P.~Haslinger, and P.~Hamilton, ``Efficient adiabatic
  spin-dependent kicks in an atom interferometer,'' {\em Physical Review
  Letters} {\bfseries 121} (2018) 040402.

\bibitem{Pagel2020lattice}
Z.~Pagel, W.~Zhong, R.~H. Parker, C.~T. Olund, N.~Y. Yao, and H.~M\"uller,
  ``Symmetric bloch oscillations of matter waves,'' {\em Physical Review A}
  {\bfseries 102} (2020) 053312.

\bibitem{purdy2017quantum}
T.~Purdy, K.~Grutter, K.~Srinivasan, and J.~Taylor, ``Quantum correlations from
  a room-temperature optomechanical cavity,'' {\em Science} {\bfseries 356}
  no.~6344, (2017) 1265--1268.

\bibitem{peskin2018introduction}
M.~Peskin and D.~Schroeder, {\em An introduction to quantum field theory}.
\newblock CRC press, 2018.

\bibitem{kadanoff1966scaling}
L.~P. Kadanoff, ``Scaling laws for {I}sing models near {$T_c$},'' {\em Physics
  Physique Fizika} {\bfseries 2} no.~6, (1966) 263.

\bibitem{wilson1971renormalization}
K.~G. Wilson, ``Renormalization group and critical phenomena. {I}.
  {R}enormalization group and the {K}adanoff scaling picture,'' {\em Physical
  review B} {\bfseries 4} no.~9, (1971) 3174.

\bibitem{weinberg1979phenomenological}
S.~Weinberg, ``Phenomenological {L}agrangians,'' {\em Physica {A}} {\bfseries
  96} no.~1-2, (1979) 327--340.

\bibitem{weinberg1995quantum}
S.~Weinberg, {\em The quantum theory of fields}, vol.~2.
\newblock Cambridge university press, 1995.

\bibitem{anastopoulos2013master}
C.~Anastopoulos and B.~Hu, ``A master equation for gravitational decoherence:
  probing the textures of spacetime,'' {\em Classical and Quantum Gravity}
  {\bfseries 30} no.~16, (2013) 165007.

\bibitem{Anastopoulos:2018drh}
C.~Anastopoulos and B.-L. Hu, ``{Comment on ''A Spin Entanglement Witness for
  Quantum Gravity'' and on ''Gravitationally Induced Entanglement between Two
  Massive Particles is Sufficient Evidence of Quantum Effects in Gravity''},''
  \href{http://arxiv.org/abs/1804.11315}{{\ttfamily arXiv:1804.11315
  [quant-ph]}}.

\bibitem{carneyprep}
D.~Carney, ``{Newton, entanglement, and the graviton},'' {\em In preparation} .

\bibitem{clerk2010introduction}
A.~A. Clerk, M.~H. Devoret, S.~M. Girvin, F.~Marquardt, and R.~J. Schoelkopf,
  ``Introduction to quantum noise, measurement, and amplification,'' {\em
  Reviews of Modern Physics} {\bfseries 82} no.~2, (2010) 1155.

\end{thebibliography}\endgroup

\end{document}